\documentclass[journal,10pt]{IEEEtran}

\IEEEoverridecommandlockouts

\usepackage{graphicx} 
\usepackage{epstopdf} 

\usepackage{cite}
\usepackage{amsmath,amssymb,amsfonts}
\usepackage{algorithmic}
\usepackage{graphicx}
\usepackage{textcomp}
\usepackage{xcolor}
\usepackage{color}

\usepackage{amssymb}
\usepackage{bbm} 
\usepackage{bm}

\usepackage{algorithm}
\usepackage{algorithmic}
\usepackage{url}
\usepackage{color}

\usepackage{graphicx} 
\usepackage{epstopdf} 

\usepackage{cite}
\usepackage{amsmath,amssymb,amsfonts}
\usepackage{algorithmic}
\usepackage{graphicx}
\usepackage{textcomp}
\usepackage{xcolor}
\usepackage{color}

\usepackage{amssymb}
\usepackage{bbm} 
\usepackage{bm}

\usepackage{algorithm}
\usepackage{algorithmic}
\usepackage{url}
\usepackage{color}

\usepackage{multirow} 

\usepackage{subfigure}

\usepackage{tablefootnote}

\usepackage{caption}

\usepackage{amsmath}
\allowdisplaybreaks[1]

\def\BibTeX{{\rm B\kern-.05em{\sc i\kern-.025em b}\kern-.08em
    T\kern-.1667em\lower.7ex\hbox{E}\kern-.125emX}}
\begin{document}

\title{Prediction, Communication, and Computing Duration Optimization for VR Video Streaming}
\author{\IEEEauthorblockN{Xing Wei, Chenyang Yang, and Shengqian Han}
\thanks{
	This work was supported by National Natural Science Foundation of China (NSFC) under Grant 61671036, Grant 61871015, and Grant 61731002.
	
	Xing Wei, Chenyang Yang, and Shengqian Han are with the School of Electronics and Information Engineering, Beihang University,  Beijing 100191, China, Shengqian Han is also with the Hangzhou Innovation Institute, Beihang University, Hangzhou 310000, China (e-mail: weixing@buaa.edu.cn, cyyang@buaa.edu.cn, sqhan@buaa.edu.cn).
	}
}

\maketitle

\vspace{-14mm}
\begin{abstract}
    Proactive tile-based video streaming can avoid motion-to-photon latency of wireless virtual reality (VR) by computing and delivering the predicted tiles to be requested before playback. All existing works either focus on designing predictors or allocating computing and communications resources. Yet  to avoid the latency, the successively executed prediction, communication, and computing tasks should be accomplished within a predetermined time. Moreover, the quality of experience (QoE) of proactive VR streaming depends on the worst performance of the three tasks. In this paper, we jointly optimize the duration of the observation window for predicting tiles and the durations for computing and transmitting the predicted tiles, aimed at balancing the performance for three tasks to maximize the QoE given arbitrary predictor and configured resources. We obtain the closed-form optimal solution by decomposing the formulated problem equivalently into two subproblems. With the optimized durations, we find a resource-limited region where the QoE increases rapidly with configured resources, and a prediction-limited region where the QoE can be improved more efficiently with a better predictor. Simulation results using three existing predictors and a real dataset validate the analysis and demonstrate the gain from the joint optimization over non-optimized counterparts.
\end{abstract}

\section{Introduction}
Wireless virtual reality (VR) can provide an immersive experience to wireless users. As the main type of VR services~\cite{3GPP_standard}, VR video usually has $360^{\circ}\times 180^{\circ}$ panoramic view with ultra high resolution (e.g. 16~K~\cite{ieeenetworkmagzine}). Evidently, delivering such video is cost-prohibitive for wireless networks. However, the range of angles that humans can see at the same time is only a limited area of the full panoramic view (about $110^{\circ}\times90^{\circ}$~\cite{FoV_size}), called field of view (FoV). This inspires  tile-based streaming\cite{adptive_tile,survey_Hsu}, which divides a full panoramic view segment into small tiles in spatial domain, and only computes and transmits the tiles overlapped with the FoVs.

To avoid dizziness in watching VR video, the motion-to-photon (MTP) latency should be low, which usually should be less than 20~ms \cite{HuaWei_whitepaper}. With reactive tile-based streaming, the tiles within the FoVs should be computed and delivered within the MTP latency after a user initiates a request, which demands for very high transmission rate and computing rate\cite{HuaWei_whitepaper,URLLC-VR}. Proactive tile-based streaming can avoid the MTP latency \cite{survey_Hsu,HuaWei_whitepaper,optimizing_VR}, which is anticipated to be the mainstream in the ultimate stage of VR video streaming\cite{HuaWei_whitepaper}.

\vspace{-0mm}\subsection{Motivation and Major Contributions}
Proactive tile-based VR video streaming contains three tasks: prediction, communication, and computing. Before the playback of a segment, the tiles to be most likely requested in the segment are first predicted using the user behavior-related data in an observation window, which are then computed and finally delivered to the user.

{To avoid the  MTP latency, these successively executed tasks should be accomplished within a pre-determined duration,} which depends on the ``coherence time'' of the user-behavior related data in watching a VR video, the employed predictor and the playback duration of a segment.

On the other hand, even without the MTP latency, the quality of experience (QoE) of a user may degrade due to the {black holes} during VR video playback ~\cite{survey_Hsu,Scalable-VR}. This occurs when the following cases happen.
(1) More tiles to be requested in a segment can be correctly predicted with longer observation window, but no enough time is left for computing and delivering all the predicted tiles before the playback of the segment. (2) More tiles can be computed and delivered with longer time for computing and communication, but some of these tiles are not requested due to the poor prediction performance. It suggests that the QoE depends on the worst performance of the three tasks.
In this context,
the performance of prediction indicates how many
tiles to be requested can be predicted correctly\cite{Fixation_Prediction,optimizing_VR}, and the performance of communication and computing refers to how many predicted tiles can be computed and delivered before playback. The prediction performance is restricted by the  ``coherence time'' even with the best predictor. Delivering and computing large amount of data within short time requires high transmission and computing rates.  To maximize the QoE with a given predictor and configured resources, the performance of the three tasks should be balanced. This can be achieved by judiciously allocating the durations for the observation window and for processing and delivering the predicted tiles.

In this paper, we strive to investigate how to match the computing rate and transmission rate to the prediction performance to maximize the QoE of proactive tile-based VR video streaming. The main contributions are summarized as follows.
\begin{itemize}
  \item To balance the performance of the three tasks, we formulate a problem to optimize the duration assigned to the observation window for any given predictor and the durations for processing and delivering the predicted tiles with given computing and transmission rates.
  \item We find the optimal durations with closed-form expressions, with a reasonable assumption on the relation between the prediction performance and the observation time, which is validated using existing predictors and a real dataset. From the optimal solution,
we figure out a single parameter to reflect the configured communication and computing resources, and find resource-limited and prediction-limited regions, which provides useful insights into the system design. In the resource-limited region, the QoE can be improved by boosting communication and computing performance. In the prediction-limited region, the QoE can be improved more effectively by enhancing the prediction performance.
\end{itemize}

\vspace{-0mm}\subsection{Related Works}
As far as the authors known, there are no prior works to co-design the prediction, communication, and computing tasks for proactive tile-based VR video streaming.

For tile prediction, a linear regression (LR) method was proposed in \cite{optimizing_VR} to predict the central point of FoV, which was then transformed into corresponding tiles~\cite{apcc}. A deep reinforcement learning technique was used in \cite{xumai_predictHM} to predict the FoV of the next frame. A deep learning method was proposed in \cite{Fixation_Prediction} that uses head mounted display (HMD) orientations, image saliency maps, and motion maps to predict the tiles to be requested. Context bandits (CB) learning technique was considered in \cite{apcc} to predict the tile requests and in \cite{CB4VR} to predict the user orientation in longitude and latitude. A sequence-to-sequence prediction method was proposed in \cite{verylong_predict} to predict the future FoV in seconds ahead.

For communication and computing resource allocation, the rendering task in computing was offloaded in \cite{edge_assisted} from HMD to multiple-access edge computing (MEC) server to reduce the bandwidth usage and computational workload on HMD. The computing and caching resources were leveraged in \cite{task_scheduling} to reduce the communication resource usage. Multicast opportunity was exploited for multiple users watching the same video in \cite{tile_multicast} to minimize the transmission energy. The uplink and downlink communication resource allocation was optimized in \cite{Mchen_uplink_downlink} to maximize a utility function that takes prediction accuracy, communication delay, and computing delay into consideration.
In \cite{YSun3C,TDang3C}, partial computing task was offloaded from the MEC server to the HMD, which reveals the tradeoff between caching and computing without considering tile prediction. In our previous work\cite{apcc}, the durations for communication and computing were optimized, while the duration for prediction was simply given.

All prior works do not investigate how to balance the performance of the three tasks in order to maximize the QoE of proactive tile-based VR streaming.

The rest of this paper is organized as follows. Section II describes the system model. Section III formulates the problem, and Section IV derives the optimal solution and provides the prediction-limited and resource-limited regions. Simulation and numerical results are provided in Section V to validate the assumption, show the QoE in the two regions, and evaluate the gain from optimizing the durations for three tasks. Section VI concludes the paper.

\section{System Model}\label{section-system_model}
Consider a proactive tile-based VR video streaming system with an MEC server co-located with a base station (BS).
Each VR video consists of $L$ segments in temporal domain, and each segment consists of $M$ tiles in spatial domain. The playback duration of each tile equals to the playback duration of a segment, denoted by $T_{\mathrm{seg}}$\cite{optimizing_VR,survey_Hsu}.
Each user is equipped with a HMD, which can measure data\footnote{The useful data for prediction include sensor-related data (i.e., the head movement orientation tracking \cite{optimizing_VR} and eye tracking data\cite{xumai_predictHM,vive} from the HMD sensors), and content-related data (i.e., the images in historical FoVs\cite{xumai_predictHM,NTHU_dataset} and  temporal-spatial saliency\cite{Fixation_Prediction,CVPR_gaze}). The audio data in a VR video are also under discussion\cite{CVPR_gaze}.
While \textit{our framework is applicable for prediction using one or multiple types of data}, we take the head movement trace as an example for easy exposition.
} of the user, send the recorded data to the MEC server, and pre-buffer segments.
While both the MEC server and the HMD can be used for rendering, we consider that the MEC server renders a video segment before delivering to the HMD.
\begin{figure}[htbp]
	\vspace{-0.4cm}
	\centering
	\begin{minipage}[t]{0.5\textwidth}
	\includegraphics[width=1\textwidth]{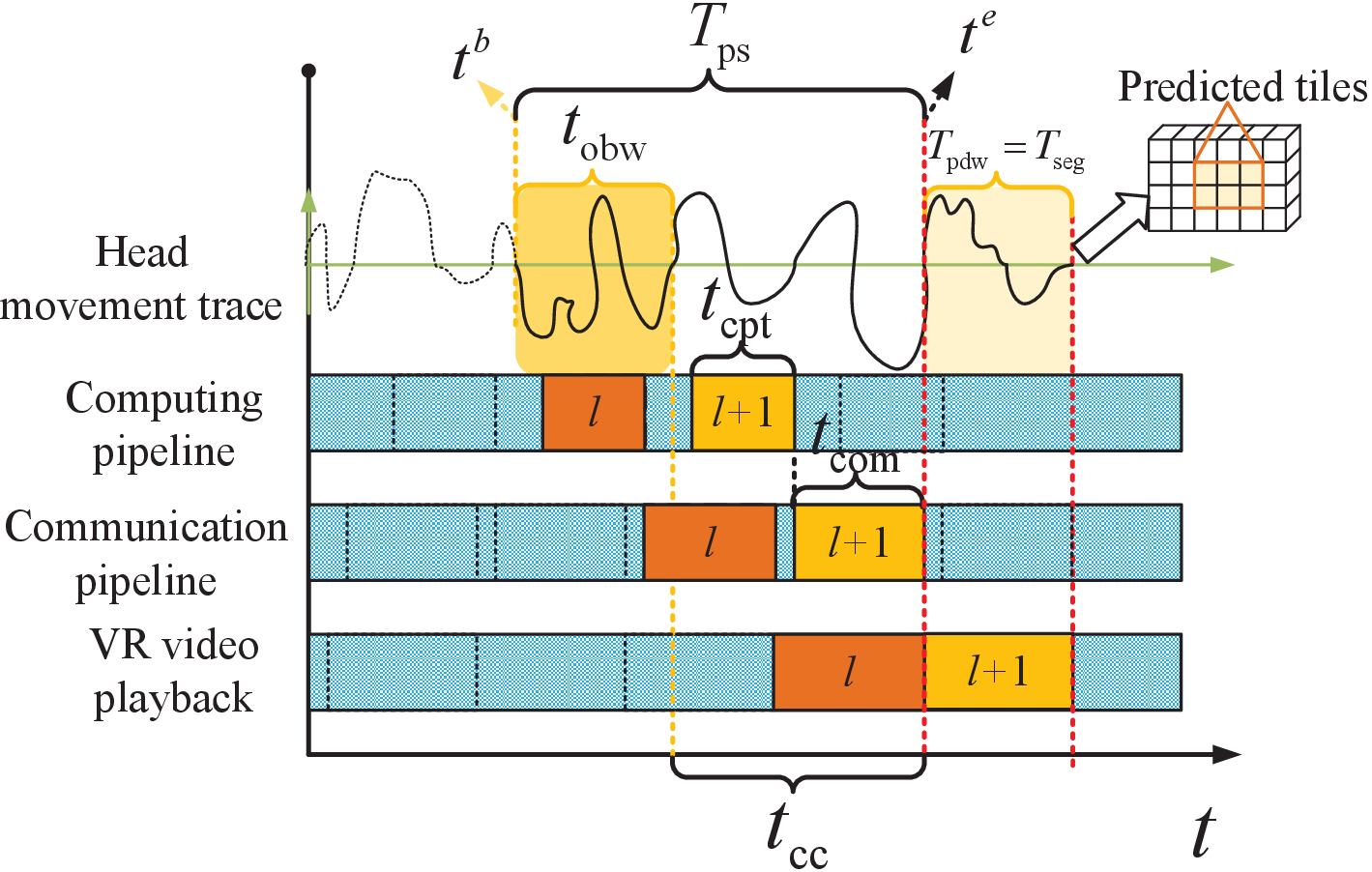}
	\end{minipage}
	\vspace{-0.2cm}
	\caption{Proactively streaming the ($l$+1)th segment of a VR video. $t^b$ is the start time for streaming the ($l$+1)th segment, and $t^e$ is the start time of playback of the ($l$+1)th segment.}
	\label{Fig:duration}
	\vspace{-0.15cm}
\end{figure} 

During the VR video playback, subsequent segments are predicted, computed, and delivered one after another\cite{HuaWei_whitepaper,survey_Hsu}, as shown in Fig. \ref{Fig:duration}.
Each segment has its own deadline for the prediction, computing, and communication, hence the three tasks for $L$ segments can be decoupled into $L$ groups of tasks. In the sequel, we take the ($l+1$)th segment as an example for elaboration.

With proactive VR streaming, the MTP latency can be avoided by processing and delivering the tiles in the ($l$+1)th segment to be requested before $t^e$.
Hence, the tiles to be played in a prediction window, which begins at $t^e$ and is with duration $T_{\mathrm{pdw}}=T_{\mathrm{seg}}$, should be predicted.

This can be accomplished by first predicting the head movement sequence
in the prediction window with the observations (say also a head movement sequence) in a time window with duration $t_{\mathrm{obw}}$, then mapping the predicted sequence into the predicted tiles~\cite{NTHU_dataset,apcc}.
Since the prediction performance for a time series depends on its auto-correlation, there exists a time instant $t^b$, before which the observations have low correlation with the sequence to be predicted and hence contribute little or even negatively for improving the prediction accuracy \cite{Predictive_xueshihou,Fixation_Prediction,head-body}. We can set $t^b$ as the start time of the observation window. At the end of the observation window, the MEC server first makes the prediction, then renders the predicted tiles with duration $t_{\mathrm{cpt}}$, and finally delivers the processed tiles with duration $t_{\mathrm{com}}$. Then, the duration between $t^b$ and $t^e$, denoted as $T_{\mathrm{ps}}$, is the total proactive streaming time available for prediction, computing, and communication, i.e., $t_{\mathrm{obw}}+t_{\mathrm{cpt}}+t_{\mathrm{com}}\leq T_{\mathrm{ps}}$.
The value of $T_\mathrm{ps}$ can be predetermined for any given dataset and predictor~\cite{Predictive_xueshihou,Fixation_Prediction,head-body}. The duration $T_\mathrm{ps}+T_{\mathrm{pdw}}$ can be regarded as the ``coherence time'' for the head movement time series.

\vspace{-0cm}
\subsection{Computing Model}
For tile-based VR video streaming, rendering contains two steps.
(1) The tiles in a segment are unpacked and the tiled-frames at the same timestamp are concatenated to generate successive two-dimensional (2D) FoVs\cite{edge_assisted}. (2) The 2D planar FoVs are converted to the three-dimensional (3D) spherical FoVs, by multiplying the image matrix of 2D FoV with a projection matrix  (this step is sometimes called ``projection'' in the rendering operation~\cite{survey_Hsu}). In practice, GPU is necessary for real-time rendering.
When the GPU with Turing architecture \cite{Turing_GPU} is used for MEC \cite{Nvidia_cloudXR,Nvidia_MEC_parameter}, the computing resource for rendering a VR video can be assigned by allocating compute unified device architecture cores \cite{Nvidia_Core} and multiple GPUs can be used at the server.

To gain useful insight, we assume that the computing resource, denoted as $\mathcal{C}_{\mathrm{total}}$ (in floating-point operations per second, FLOPS), is equally allocated among $K$ users. Then,  the number of bits that can be rendered per second, referred to as the \textit{computing rate} for the $k$th user, is
\begin{align}\label{C_cpt_final}
C_{\mathrm{cpt},k} \triangleq \frac{\mathcal{C}_{r,k}}{\mu_r} (\textit{in bits/s}),
\end{align}
where $\mathcal{C}_{r,k}=\mathcal{C}_{\mathrm{total}}/K$ is the computing resource assigned to the $k$th user for rendering, and
\begin{align}
\mu_r = \frac{T_r \mathcal{C}_{r}}{\gamma_{\textit{fov}} R_w R_h b}\nonumber
\end{align}
is the required floating-point operations (FLOPs) of rendering one bit of FoV \textit{in FLOPs/bit},
$T_r$ is the time used for rendering a FoV,
$\gamma_{\textit{fov}}$ is the ratio of FoV in a frame,
$R_w$ and $R_h$ are respectively the pixels in wide and high of a frame, and $b$ is the number of bits per pixel relevant to color depth\cite{HuaWei_Cloud_VR}.


\vspace{-0cm}\subsection{Transmission Model}
The BS equipped with $N_t$ antennas serves $K$ single-antenna users with zero-forcing beamforming.
The instantaneous transmission rate at the $i$th time slot for the $k$th user is
\begin{align}\label{C_com_inst}
C_{\mathrm{com},k}^{i}=B\log_2 \left(1+\frac{p^i_k d_k^{-\alpha}|\tilde{h}^i_k|^2}{\sigma^2} \right),
\end{align}
where $B$ is the bandwidth, $\tilde{h}^i_k\triangleq (\mathbf{h}^i_k)^H\mathbf{w}^i_k$ is the equivalent channel gain, $p^i_k$ and $\mathbf{w}^i_k$ are respectively the transmit power and beamforming vector for the $k$th user,  $d_k$ and $\mathbf{h}^i_k\in\mathbb{C}^{N_t}$ are respectively the distance and the small scale channel vector from the BS to the $k$th user, $\alpha$ is the path-loss exponent, $\sigma^2$ is the noise power, and $(\cdot)^{H}$ denotes conjugate transpose.

We consider indoor users as in the literature of wireless VR video streaming, where the distances of users usually change slightly \cite{LSTM_update,NTHU_dataset,Proc_IEEE} and are assumed fixed. We consider time-varying small-scale channels, which are assumed as remaining constant in each time slot with duration $\Delta T$ and changing among time slots independently with identical distribution.
With the proactive transmission, the predicted tiles in a segment should be transmitted with duration $t_{\mathrm{com}}$. The number of bits transmitted within $t_{\mathrm{com}}$ is $\overline{C}_{\mathrm{com},k}t_{\mathrm{com}}$, where $\overline{C}_{\mathrm{com},k}$ is the time-average rate in $t_{\mathrm{com}}$. When $t_{\mathrm{com}} \gg \Delta T$, $\overline{C}_{\mathrm{com},k}=\frac{1}{N_s} \sum_{i=1}^{N_s}C_{\mathrm{com},k}^{i}$, where $N_s$ is the number of time slots in $t_{\mathrm{com}}$.
Since future channel is unknown when making the optimization, we use ensemble-average rate $E_h\{C_{\mathrm{com},k}\}$ to approximate time-average rate $\overline{C}_{\mathrm{com},k}$, where $E_h\{\cdot\}$ is the expectation over $h$, which is very accurate when $N_s$ or $N_t/K$ is large as validated via extensive simulations.
To ensure fairness among users in terms of QoE, the transmit power is allocated to compensate the path loss, i.e., ${p}^i_k=\frac{\beta}{d_k^{-\alpha}}$, where $\beta$ can be obtained from $\beta(\sum_{k=1}^{K}\frac{1}{d_k^{-\alpha}})={P}$ and ${P}$ is the maximal transmit power of the BS. Then, the ensemble-average transmission rate for each user is equal.

In the sequel, we first consider one user for analysis and then show the impact of $K$. We use $C_{\mathrm{com}}$ to replace $E_h\{C_{\mathrm{com},k}^{i}\}$ and use $C_{\mathrm{cpt}}$ to replace $C_{{\mathrm{cpt}},k}$ for notational simplicity.

\vspace{-0cm}\section{Problem Formulation for Optimizing the Durations for Three Tasks}
In this section, we first introduce the performance metrics for the tile prediction task, for the computing and communication tasks, and for the QoE in watching a VR video. To balance the performance of the three tasks for maximizing the QoE, we then formulate a problem to jointly optimize the duration of the observation window with any given predictor, and the durations for communication and computing with given computing rate and transmission rate.

 \vspace{-0cm}\subsection{Performance Metric of Tile Prediction}\label{section:DoO_def}
Degree of overlap (DoO) has been used to measure the overlap between the predicted and the requested frames of a panoramic video \cite{xumai_predictHM,LSTM_update}. A larger value of DoO indicates a better prediction.
To reflect the prediction performance for proactive tile-based streaming, we consider \textit{segment DoO (seg-DoO)}, which measures the overlap of the predicted tiles and the requested tiles in a segment. From this perspective, the DoO used in \cite{xumai_predictHM,LSTM_update} can be considered as a special case of seg-DoO where a segment contains only one frame\cite{LSTM_update}.
The seg-DoO of the $l$th segment is defined as 
\begin{align}
\mathrm{DoO}_{l}^{\mathrm{seg}}(t_{\mathrm{obw}}) \triangleq \frac{\mathbf{q}_l^\mathsf{T}\cdot\mathbf{p}_l({t_{\mathrm{obw}}})  }{\|\mathbf{q}_l\|_1},\nonumber
\end{align}
where $\mathbf{q}_{l}\triangleq [q_{l,1},...,q_{l,M}]^\mathsf{T}$ denotes the ground-truth of the tile requests for the segment with $q_{l,m}\in\{0,1\}$, $\mathbf{e}_{l}({t_{\mathrm{obw}}})\triangleq [e_{l,1}({t_{\mathrm{obw}}}),...,e_{l,M}({t_{\mathrm{obw}}})]^\mathsf{T}$ denotes the predicted tile requests for the segment with $e_{l,m}({t_{\mathrm{obw}}})\in\{0,1\}$,
$(\cdot)^\mathsf{T}$ denotes transpose of a vector, and $\|\cdot\|_1$ denotes the $\ell_1$ norm of a vector. When the $m$th tile in the $l$th segment is truly requested, $q_{l,m}=1$, otherwise $q_{l,m}=0$. When the tile is predicted to be requested, $e_{l,m}({t_{\mathrm{obw}}})=1$, otherwise it is zero.

We use average seg-DoO to measure the prediction performance for a VR video, which is
\begin{align}
  \mathcal{D}(t_{\mathrm{obw}}) \triangleq \frac{1}{L}\sum_{l=1}^L\mathrm{DoO}_{l}^{\mathrm{seg}}(t_{\mathrm{obw}})
  = \sum_{n=0}^{\infty} a_n t_{\mathrm{obw}}^n, \label{DoO_power_series}
\end{align}
where the second equality is the power series expansion, which holds for any infinitely differentiable function. The value of $a_n$ depends on the predictor.

For any pre-determined value of $T_{\mathrm{ps}}$, a predictor can be more accurate with a longer observation window, because the tiles to be predicted are closer to and hence are more correlated with the head movement sequence having been observed, as shown in Fig. \ref{Fig:duration}. This gives rise to a reasonable assumption as follows.

\textbf{Assumption 1}: $\mathcal{D}(t_{\mathrm{obw}})$ is a monotonically increasing function of $t_{\mathrm{obw}}$.


\vspace{-0mm}
\subsection{Completion Rate of Communication and Computing Tasks}\label{subsection:CC_capacity}
To reflect the performance of the system for rendering and delivering the predicted tiles in a segment, we define the completion rate of communication and computing (CC) tasks as 
\begin{align}
  S_{\mathrm{cc}}(t_{\mathrm{com}},t_{\mathrm{cpt}}) &\triangleq \frac{\min\{S(t_{\mathrm{com}},t_{\mathrm{cpt}}),N\}}{N}, \label{S-cc}
\end{align}
where $N\triangleq \|\mathbf{e}_l({t_{\mathrm{obw}}})\|_1$ is the number of predicted tiles in a segment, $S(t_{\mathrm{com}},t_{\mathrm{cpt}})$ is the number of tiles in the segment that can be delivered with $t_{\mathrm{com}}$  and computed with  $t_{\mathrm{cpt}}$, which is
\begin{align}
S(t_{\mathrm{com}},t_{\mathrm{cpt}}) \triangleq \min\left\{\frac{C_{\mathrm{com}}t_{\mathrm{com}}}{s_{\mathrm{com}}^{}}, \frac{C_{\mathrm{cpt}}t_{\mathrm{cpt}}}{s_{\mathrm{cpt}}^{}}\right\}, \label{S_def}
\end{align}
where $s_{\mathrm{com}} = r_w r_h b  N_{\textit{tf}}/\gamma_{c}$ \cite{HuaWei_Cloud_VR} is the number of bits in each tile for transmission, $s_{\mathrm{cpt}} = r_w r_h bN_{\textit{tf}}$ is the number of bits in a tile for rendering, $r_w$ and $r_h$ are the pixels in wide and high of a tiled-frame, $\gamma_c$ is the compression ratio, and $N_{\textit{tf}}$ is the number of tiled-frame in a tile.

After substituting \eqref{S_def} into \eqref{S-cc}, the completion rate of CC tasks can be expressed as
\begin{align}\label{S_CC_def_final}
  S_{\mathrm{cc}}(t_{\mathrm{com}},t_{\mathrm{cpt}}) =  \min\left\{\frac{C_{\mathrm{com}}t_{\mathrm{com}}}{s_{\mathrm{com}}N}, \frac{C_{\mathrm{cpt}}t_{\mathrm{cpt}}}{s_{\mathrm{cpt}}N},1\right\}.
\end{align}

\subsection{Metric of Quality of Experience}
For proactive tile-based streaming, there will be no MTP latency if the constraint
$t_{\mathrm{obw}} + t_{\mathrm{com}} + t_{\mathrm{cpt}} \leq T_{\mathrm{ps}}$ can be satisfied.
Yet black holes will appear if either the requested tiles cannot be predicted correctly, or some predicted tiles cannot be computed and delivered before playback.
To reflect the percentage of the correctly predicted tiles for any given predictor with observation window of duration $t_{\mathrm{tow}}$ that can be computed and delivered within durations $t_{\mathrm{com}}
$ and $t_{\mathrm{cpt}}$ among all the requested tiles, we consider the following QoE metric
\begin{align}
\mathrm{QoE}&\triangleq \frac{1}{L}\sum_{l=1}^{L}\frac{\mathbf{q}_l^\mathsf{T}\cdot\big(\mathbf{e}_l({t_{\mathrm{obw}}})\cdot S_{\mathrm{cc}}(t_{\mathrm{com}},t_{\mathrm{cpt}})\big)  }{\|\mathbf{q}_l\|_1}\nonumber\\
&=\Big(\frac{1}{L}\sum_{l=1}^{L}\frac{\mathbf{q}_l^\mathsf{T}\cdot\mathbf{e}_l({t_{\mathrm{obw}}})  }{\|\mathbf{q}_l\|_1}\Big)\cdot S_{\mathrm{cc}}(t_{\mathrm{com}},t_{\mathrm{cpt}})\nonumber\\
&=\mathcal{D}(t_{\mathrm{obw}})\cdot S_{\mathrm{cc}}(t_{\mathrm{com}},t_{\mathrm{cpt}}),\label{QoE-Metric}
\end{align}
where $\mathbf{e}_l(t_{\mathrm{obw}})\cdot S_{\mathrm{cc}}(t_{\mathrm{com}},t_{\mathrm{cpt}})$
is the predicted tiles that can be delivered and computed with durations $t_{\mathrm{com}}$ and $t_{\mathrm{cpt}}$. When the value of the QoE is $100\%$, all the requested tiles in a VR video are proactively computed and delivered before playback.
The QoE can be increased by improving the prediction performance or the computing and communication performance.
To reduce the impact of wrong prediction, one can proactively streaming extra tiles in addition to the predicted tiles. For example, we can set $N=\left \lceil{\gamma_{\mathrm{pt}}M}\right \rceil $ \cite{HuaWei_Cloud_VR}, where $\gamma_{\mathrm{pt}}=\gamma_{\textit{fov}}+\gamma_{\textit{extra}}\in[0,1]$, $\gamma_{\textit{fov}}$ and $\gamma_{\textit{extra}}$ are respectively the percentage of the tiles in a FoV in a frame and the percentage of the tiles additionally computed and transmitted \cite{HuaWei_Cloud_VR}, and $\left \lceil{\cdot}\right \rceil$ is the ceil function.


\vspace{-4mm}
\subsection{Joint Optimization of the Durations for Prediction, Computing, and Communications}
The duration optimization problem to maximize the QoE with given computing rate $C_{\mathrm{cpt}}$ and transmission rate $C_{\mathrm{com}}$ can be formulated as
\vspace{-0.02cm}
\begin{subequations}
\label{P1-all}
\begin{align}
  \textbf{P1}:& \max_{t_{\mathrm{obw}},t_{\mathrm{com}},t_{\mathrm{cpt}}} \  \ \mathcal{D}(t_{\mathrm{obw}})\cdot S_{\mathrm{cc}}(t_{\mathrm{com}},t_{\mathrm{cpt}}) \label{P1_obj}\\
  &  \ \ \ \  \  s.t.  \ \  \ \ \ \ \  t_{\mathrm{obw}} + t_{\mathrm{com}} + t_{\mathrm{cpt}} = T_{\mathrm{ps}},\label{P1_CCO}\\
  & \ \ \ \ \  \  \ \  \ \ \ \ \ \ \ \ t_{\mathrm{obw}} \geq \tau,\ \  t_{\mathrm{com}}, t_{\mathrm{cpt}} \geq 0, \label{P1_duration}
\end{align}
\end{subequations}
where $\tau$ is the minimal duration of the observation window, which depends on the specific prediction method\cite{apcc,LSTM_update}.
In this problem, we replace the constraint imposed by ensuring zero MTP latency with \eqref{P1_CCO}, because the objective is a monotonically  increasing function of $t_{\mathrm{obw}}$, $t_{\mathrm{cc}}$, and $t_{\mathrm{cpt}}$ under \textbf{Assumption 1}.

\vspace{-0mm}\section{Optimal Durations and Prediction-limited and Resource-limited Regions}
In this section, we investigate how to balance the performance of the three tasks by solving the joint duration optimization problem. To this end, we first decouple problem \textbf{P1} equivalently into two subproblems, thanks to the separability of the objective function in (\ref{P1_obj}) with respect to $t_{\mathrm{com}}$, $t_{\mathrm{cpt}}$ and $t_{\mathrm{obw}}$, and obtain  the global optimal solution with closed-form expression. Then, we show that the system may operate in a prediction-limited or a resource-limited region, depending on the computing rate and transmission rate.

\vspace{-0mm}\subsection{Problem Decomposition}
First, we optimize $t_{\mathrm{com}}$ and $t_{\mathrm{cpt}}$ for arbitrarily given total time for computing and delivering the predicted tiles $t_{\mathrm{cc}} \triangleq t_{\mathrm{com}} +t_{\mathrm{cpt}}$ that satisfies \eqref{P1_CCO} from the following problem
\begin{subequations}
\begin{align}
  \textbf{P2}: & \max_{t_{\mathrm{com}},t_{\mathrm{cpt}}} \ \ \   S_{\mathrm{cc}}(t_{\mathrm{com}},t_{\mathrm{cpt}})\label{P2_obj}\\
  & \ \ \   s.t.\ \ \ \ \   t_{\mathrm{com}} + t_{\mathrm{cpt}} = t_{\mathrm{cc}},\\ 
  & \ \ \ \ \ \ \ \ \ \ \ \   t_{\mathrm{com}}, t_{\mathrm{cpt}} \geq 0.
\end{align}
\end{subequations}
Then, we optimize $t_{\mathrm{obw}}$ and $t_{\mathrm{cc}}$ from the following problem
\begin{subequations}
\begin{align}
  \textbf{P3}:& \max_{t_{\mathrm{obw}},t_{\mathrm{cc}}} \  \ \ \mathcal{D}(t_{\mathrm{obw}})\cdot S^{*}_{\mathrm{cc}}(t_{\mathrm{cc}}) \label{P3_obj}\\
  &  \ \ \  s.t.  \ \ \ \ \   t_{\mathrm{obw}} + t_{\mathrm{cc}}  = T_{\mathrm{ps}}\label{P3_S_CC},\\
  & \ \ \ \ \ \ \ \ \ \ \ \ t_{\mathrm{obw}} \geq \tau,\ \  \label{P3_duration}
\end{align}
\end{subequations}
where $S^{*}_{\mathrm{cc}}(t_{\mathrm{cc}})$ is the maximized objective of problem \textbf{P2}
as a function of $t_{\mathrm{cc}}$.


Problem \textbf{P2} optimizes the durations for communication and computing with given values of $C_{\mathrm{cpt}}$ and $C_{\mathrm{com}}$, so as to maximize the term of completion rate of CC tasks in \eqref{P1_obj}. The solution of problem \textbf{P3} can match the maximal completion rate of CC tasks to the prediction performance of any given predictor, so as to maximize the QoE in \eqref{P1_obj}.

\vspace{-0mm}\subsection{Solution of Problem \textbf{P2}}
By eliminating $t_{\mathrm{com}} = t_{\mathrm{cc}} - t_{\mathrm{cpt}}$ and considering the expression of
$S_{\mathrm{cc}}(t_{\mathrm{com}},t_{\mathrm{cpt}})$ in \eqref{S_CC_def_final}, problem \textbf{P2} can be equivalently transformed as
\vspace{-0.01cm}
\begin{subequations}\label{P3}
\begin{align}
&  \max_{t_{\mathrm{cpt}},S_{\mathrm{cc}}(t_{\mathrm{cc}})} \ \ \ \ S_{\mathrm{cc}}(t_{\mathrm{cc}}) \\
&   \ \ \  \ \ \ \  s.t.  \ \ \  \ \ \  S_{\mathrm{cc}}(t_{\mathrm{cc}}) \leq \frac{C_{\mathrm{com}}(t_{\mathrm{cc}} - t_{\mathrm{cpt}})}{s_{\mathrm{com}}N},\label{P3_C1}\\
&       \ \  \ \  \ \ \ \  \ \  \ \ \ \  \  \ \
S_{\mathrm{cc}}(t_{\mathrm{cc}}) \leq \frac{C_{\mathrm{cpt}}t_{\mathrm{cpt}}}{s_{\mathrm{cpt}}N},\label{P3_C2}\\
&     \ \  \ \  \ \ \ \  \ \  \ \ \ \  \  \ \
S_{\mathrm{cc}}(t_{\mathrm{cc}}) \leq 1.  \label{P3_C3}
\end{align}
\end{subequations}
We do not express $t_{\mathrm{cpt}}$ as a function of $t_{\mathrm{cc}}$ in this problem for notational simplicity.
As derived in Appendix \ref{Proof:MEC_t_com_t_cpt_KKT}, the solution of problem \eqref{P3} is,
\begin{subequations}\label{P3:opt_solution_general}
\begin{align}
&t_{\mathrm{cpt}}^{*}(t_{\mathrm{cc}}) =
\left
\{\begin{array}{lr}
\frac{C_{\mathrm{com}}s_{\mathrm{cpt}}}{C_{\mathrm{com}}s_{\mathrm{cpt}} + C_{\mathrm{cpt}}s_{\mathrm{com}}} t_{\mathrm{cc}},& t_{\mathrm{cc}} \leq T_{\mathrm{cc}}^{\max},  \\
\alpha,\alpha\in(\frac{s_{\mathrm{cpt}}N}{C_{\mathrm{cpt}} },\infty), & t_{\mathrm{cc}} \geq T_{\mathrm{cc}}^{\max},\\
\end{array}
\right.\label{P3:opt_solution_general_t_cpt}\\
& S_{\mathrm{cc}}^{*}(t_{\mathrm{cc}}) =
\left
\{\begin{array}{lr}
\frac{t_{\mathrm{cc}}}{T_{\mathrm{cc}}^{\max}} ,& t_{\mathrm{cc}} \leq T_{\mathrm{cc}}^{\max},  \\
1,& t_{\mathrm{cc}} \geq T_{\mathrm{cc}}^{\max},
\end{array}
\right.\label{P3:opt_solution_general_S_CC}
\end{align}
\end{subequations}
where $T_{\mathrm{cc}}^{\max}$ is the duration to deliver and compute the predicted tiles in a segment with expression
\begin{align}\label{T_CC_max}
T_{\mathrm{cc}}^{\max} \triangleq \frac{s_{\mathrm{com}}N}{C_{\mathrm{com}}} + \frac{s_{\mathrm{cpt}}N}{C_{\mathrm{cpt}}}.
\end{align}
The value of $1/{T_{\mathrm{cc}}^{\max}}$  monotonically increases with transmission rate or computing rate of a VR user, which reflects the tradeoff between communication and computing.

From \eqref{P3:opt_solution_general_t_cpt} and the definition of $t_{\mathrm{cc}}$, we have
\begin{align}
&t_{\mathrm{com}}^{*}(t_{\mathrm{cc}}) =
\frac{C_{\mathrm{cpt}}s_{\mathrm{com}}}{C_{\mathrm{com}}s_{\mathrm{cpt}} + C_{\mathrm{cpt}}s_{\mathrm{com}}} t_{\mathrm{cc}}.
\label{P3:opt_solution_general_t_com}
\end{align}

%

\subsection{Solution of Problem \textbf{P3}}
We can observe from \eqref{P3:opt_solution_general_S_CC} that $S_{\mathrm{cc}}^{*}(t_{\mathrm{cc}})$ first increases with $t_{\mathrm{cc}}$ until $t_{\mathrm{cc}}$ reaches $T_{\mathrm{cc}}^{\max}$, after which further increasing $t_{\mathrm{cc}}$ does not improve $S_{\mathrm{cc}}^{*}(t_{\mathrm{cc}})$. On the other hand, increasing $t_{\mathrm{cc}}$ decreases $t_{\mathrm{obw}}$ as shown in \eqref{P3_S_CC}, which leads to the reduction of $\mathcal{D}(t_{\mathrm{obw}})$ according to \textbf{Assumption 1}. This suggests that the optimal value of $t_{\mathrm{cc}}$ must satisfy the constraint $t_{\mathrm{cc}}\leq T_{\mathrm{cc}}^{\max}$.

By substituting $S_{\mathrm{cc}}^{*}(t_{\mathrm{cc}})$ under $t_{\mathrm{cc}} \leq T_{\mathrm{cc}}^{\max}$ case in (\ref{P3:opt_solution_general_S_CC}), problem \textbf{P3} can be re-written as
\begin{subequations}\label{P4}
\begin{align}
   &  \max_{t_{\mathrm{obw}},t_{\mathrm{cc}}} \   \ \   \mathcal{D}(t_{\mathrm{obw}}) \cdot\frac{t_{\mathrm{cc}}}{T_{\mathrm{cc}}^{\max}} \label{P4_obj}\\
   & \ \ \ s.t.\ \ \ \  t_{\mathrm{obw}}+ t_{\mathrm{cc}} = T_{\mathrm{ps}},  \label{P4_leq} \\
    & \ \ \ \ \ \ \ \ \ \ \ t_{\mathrm{obw}} \geq \tau,\\
    &  \ \ \ \ \  \ \ \  \  \ \ 0 \leq t_{\mathrm{cc}} \leq T_{\mathrm{cc}}^{\max}.
\end{align}
\end{subequations}
As derived in Appendix \ref{Proof:MEC_t_obw_KKT}, the solution of problem \eqref{P4} is
\begin{subequations}
\begin{align}
&t_{\mathrm{obw}}^{*}=
\left
\{\begin{array}{ll}
H ,&  \phi(H) > 0 \ \textrm{and}\ \Phi = \emptyset,  \\
T_{\Phi}^*,   & \phi(H) = 0 \ \textrm{and}\ \Phi \neq \emptyset,\\
H ,&  \phi(H) > 0,  \Phi \neq \emptyset \ \textrm{and}\ f(H) \geq f(T_{\Phi}^*),\\
T_{\Phi}^*,   & \phi(H) > 0,  \Phi \neq \emptyset \ \textrm{and}\ f(H) < f(T_{\Phi}^*),
\end{array}
\right.\label{opt_t_obw_general_case}\\
&t_{\mathrm{cc}}^*=
\left
\{\begin{array}{ll}
 W ,&  \phi(H) > 0 \ \textrm{and}\ \Phi = \emptyset,  \\
T_{\mathrm{ps}} - T_{\Phi}^*,   & \phi(H) = 0 \ \textrm{and}\ \Phi \neq \emptyset, \\
W ,&  \phi(H) > 0,  \Phi \neq \emptyset \ \textrm{and}\ f(H) \geq f(T_{\Phi}^*),\\
T_{\mathrm{ps}} - T_{\Phi}^*,   & \phi(H) > 0,  \Phi \neq \emptyset \ \textrm{and}\ f(H) < f(T_{\Phi}^*),
\end{array}
\right.\label{opt_CC_general_case}
\end{align}
\end{subequations}
where $H \triangleq \max\{T_{\mathrm{ps}} - T_{\mathrm{cc}}^{\max},\tau\}$, $W \triangleq \min\{T_{\mathrm{cc}}^{\max}, T_{\mathrm{ps}} - \tau \}$, $T_{\Phi}^{*} = \mathop{\mathrm{argmax}}\limits_{t_{\mathrm{obw}}} f(t_{\mathrm{obw}}), t_{\mathrm{obw}}\in\Phi$, $f(x)\triangleq \mathcal{D}(x)\cdot\frac{(T_{\mathrm{ps}} - x )}{T_{\mathrm{cc}}^{\max}}$, $\emptyset$ denotes an empty set, and
\begin{subequations}\label{H_W}
\begin{align}
&\phi(t_{\mathrm{obw}})\triangleq \mathcal{D}(t_{\mathrm{obw}}) - \frac{\mathrm{d} \mathcal{D}(t_{\mathrm{obw}})}{\mathrm{d} t_{\mathrm{obw}}}\cdot(T_{\mathrm{ps}} - t_{\mathrm{obw}})\nonumber\\
&\ \ \ \ \ \ \ \ \ = \sum_{n=0}^{\infty} a_n t_{\mathrm{obw}}^n - (\sum_{n=1}^{\infty} n a_n t_{\mathrm{obw}}^{n-1})(T_{\mathrm{ps}} - t_{\mathrm{obw}}),\label{phi_def}\\
&\Phi\triangleq\big\{t_{\mathrm{obw}}|\phi(t_{\mathrm{obw}}) = 0, t_{\mathrm{obw}} \geq \max\{T_{\mathrm{ps}} - T_{\mathrm{cc}}^{\max}, \tau \}\big\}.\label{obj_defe}
\end{align}
\end{subequations}


\vspace{-0mm}\subsection{Solution of Problem \textbf{P1}}
The optimal durations for communication and computing can be obtained by substituting (\ref{opt_CC_general_case}) into (\ref{P3:opt_solution_general_t_com}) and (\ref{P3:opt_solution_general_t_cpt}).
The optimal duration of the observation window is \eqref{opt_t_obw_general_case}.
By substituting (\ref{opt_CC_general_case}) into \eqref{P3:opt_solution_general_S_CC}, we obtain the maximal achievable completion rate of CC tasks with $t_{\mathrm{cc}}^*$ as

\vspace{-2mm}\begin{align}
S_{\mathrm{cc}}^*(t_{\mathrm{cc}}^*) =
\left
\{\begin{array}{ll}
\frac{W}{T_{\mathrm{cc}}^{\max}}  ,&  \phi(H) > 0 \ \textrm{and}\ \Phi = \emptyset,  \\
\frac{T_{\mathrm{ps}} - T_{\Phi}^*}{T_{\mathrm{cc}}^{\max}} ,   & \phi(H) = 0 \ \textrm{and}\ \Phi \neq \emptyset, \\
\frac{W}{T_{\mathrm{cc}}^{\max}}  ,&  \phi(H) > 0,  \Phi \neq \emptyset \ \textrm{and}\ f(H) \geq f(T_{\Phi}^*),\\
\frac{T_{\mathrm{ps}} - T_{\Phi}^*}{T_{\mathrm{cc}}^{\max}} ,   & \phi(H) > 0,  \Phi \neq \emptyset \ \textrm{and}\ f(H) < f(T_{\Phi}^*).
\end{array}
\right.\label{opt_S_CC_general_case}
\end{align}

\vspace{-0mm}\subsection{Resource-limited Region and Prediction-limited Region}\label{number_results}
In this subsection, we show that the system may operate in a resource-limited region or a prediction-limited region.
We consider the optimal durations under the case where $\phi(H) > 0 \ \textrm{and}\ \Phi = \emptyset$, which can be obtained from \eqref{opt_t_obw_general_case} and \eqref{opt_CC_general_case} as
\begin{subequations}\label{opt_slt_simple}
\begin{align}
&t_{\mathrm{obw}}^* =
\left
\{\begin{array}{lr}
\tau,& T_{\mathrm{cc}}^{\max}> T_{\mathrm{ps}} - \tau, \\
T_{\mathrm{cc}}^{\max} - \tau, & T_{\mathrm{cc}}^{\max}< T_{\mathrm{ps}} - \tau,\\
\end{array}
\right.\label{opt_slt_simplea} \\
& t_{\mathrm{cc}}^* =
\left
\{\begin{array}{lr}
T_{\mathrm{ps}} - \tau, & T_{\mathrm{cc}}^{\max}> T_{\mathrm{ps}} - \tau,  \\
T_{\mathrm{cc}}^{\max}, & T_{\mathrm{cc}}^{\max}< T_{\mathrm{ps}} - \tau,
\end{array}
\right.\label{opt_slt_simpleb}
\end{align}
\end{subequations}
because only this case yields feasible solution for the three prediction methods and the real dataset to be considered in the next section.
Both the relations $t_{\mathrm{obw}}^{*}\sim\frac{1}{T_{\mathrm{cc}}^{\max}}$ in \eqref{opt_slt_simplea} and $t_{\mathrm{cc}}^*\sim\frac{1}{T_{\mathrm{cc}}^{\max}}$ in \eqref{opt_slt_simpleb} can be divided into two regions with a boundary line $T_{\mathrm{cc}}^{\max} = T_{\mathrm{ps}} - \tau$.

When the value of $\frac{1}{T_{\mathrm{cc}}^{\max}}$ is small, i.e., at least one type of the communication and computing resources is limited, not all the predicted tiles can be delivered or computed. To increase the completion rate of CC tasks,
all remaining time should be used for computing or delivering after satisfying the minimal duration required by the observation window, i.e., $t_{\mathrm{obw}} = \tau$. We refer to this region as \emph{``Resource-limited region"}, where $T_{\mathrm{cc}}^{\max}> T_{\mathrm{ps}} - \tau$.
When the value of $\frac{1}{T_{\mathrm{cc}}^{\max}}$ approaches $T_{\mathrm{cc}}^{\max} = T_{\mathrm{ps}} - \tau$ and  further increases, the completion rate of CC tasks matches the prediction performance, i.e., all the predicted tiles can be delivered and computed.
Yet as the value of $\frac{1}{T_{\mathrm{cc}}^{\max}}$ exceeds such a boundary, the prediction performance becomes the bottleneck of improving QoE. Since the prediction performance can be improved by increasing $t_{\mathrm{obw}}$, the optimal solution allocates all the remaining time for the observation window, i.e., $t_{\mathrm{obw}}^{*} = T_{\mathrm{ps}} - T_{\mathrm{cc}}^{\max}$.
We refer to this region with $T_{\mathrm{cc}}^{\max}< T_{\mathrm{ps}} - \tau$ as \emph{``Prediction-limited region"}.

\begin{figure}[htbp]
\vspace{-0.5cm}
    \centering
    \subfigure[$\frac{1}{T_{\mathrm{cc}}^{\max}}$ $\sim C_{\mathrm{com}}$  and $C_{\mathrm{cpt}}$]{\label{Fig:opt_slt_t_CC}
        \begin{minipage}[c]{0.75\linewidth}
            \centering
            \includegraphics[width=1\textwidth]{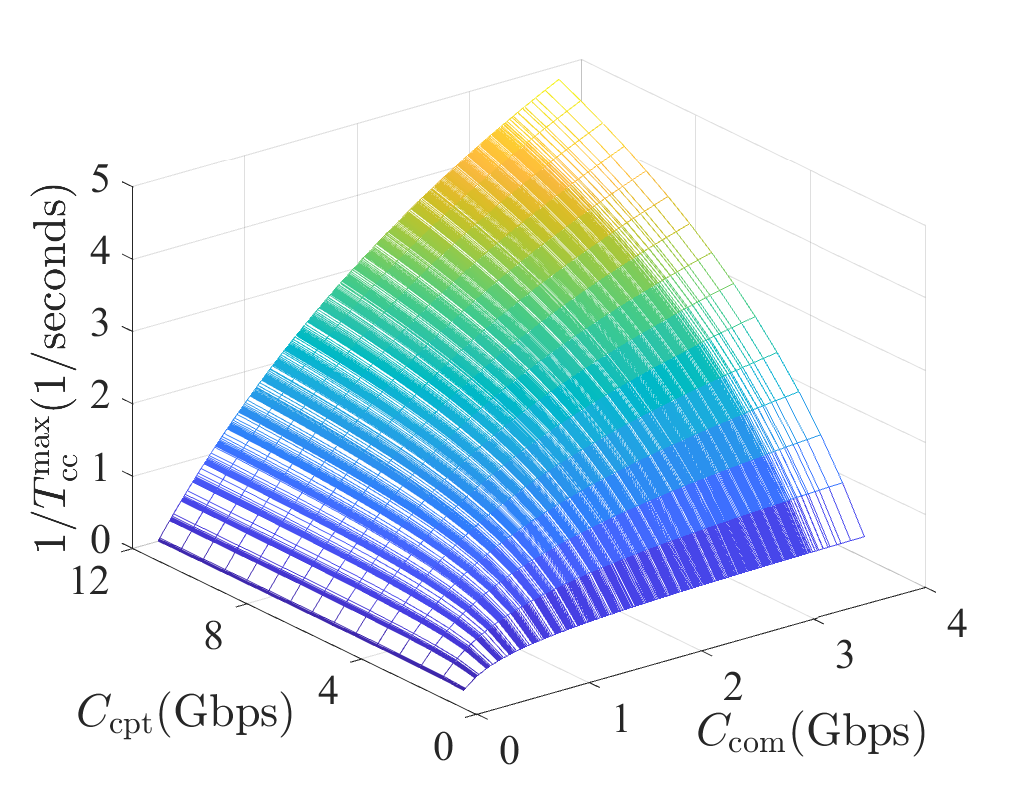}
        \end{minipage}
    }
    \subfigure[$t_{\mathrm{obw}}^{*}$ and $t_{\mathrm{cc}}^*$ $\sim \frac{1}{T_{\mathrm{cc}}^{\max}}$]{\label{Fig:opt_slt_2D_example}
        \begin{minipage}[c]{0.75\linewidth}
            \centering
            \includegraphics[width=1\textwidth]{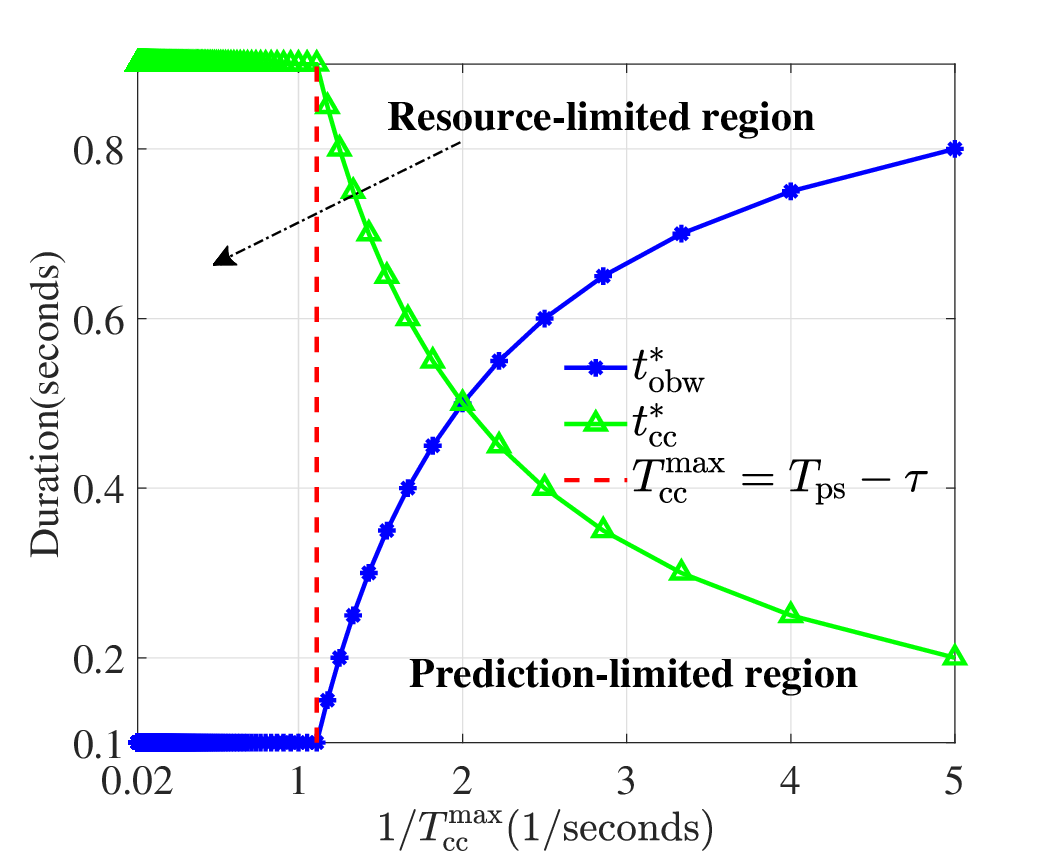}
        \end{minipage}
    }
        \vspace{-0.5cm}
    \caption{The relation of $\frac{1}{T_{\mathrm{cc}}^{\max}}$ with $C_{\mathrm{com}}$  and $C_{\mathrm{cpt}}$, and the optimal durations v.s. $\frac{1}{T_{\mathrm{cc}}^{\max}}$.}\label{Fig:T_cc_map_CC}	
    \vspace{-0.3cm}
\end{figure}

To help understand the relation of the parameter $\frac{1}{T_{\mathrm{cc}}^{\max}}$ with $C_{\mathrm{com}}$  and $C_{\mathrm{cpt}}$, we provide the values of $\frac{1}{T_{\mathrm{cc}}^{\max}}$ obtained from \eqref{T_CC_max} given different computing rate and transmission rate in Fig. \ref{Fig:opt_slt_t_CC}, where we set $N=8$, $s_{\mathrm{com}} = 52$ Mbits, and $s_{\mathrm{cpt}} = 124$ Mbits (to be explained later).

\begin{figure}[htbp]
\vspace{-0.5cm}
    \centering
        \begin{minipage}[c]{0.75\linewidth}
            \centering
            \includegraphics[width=1\textwidth]{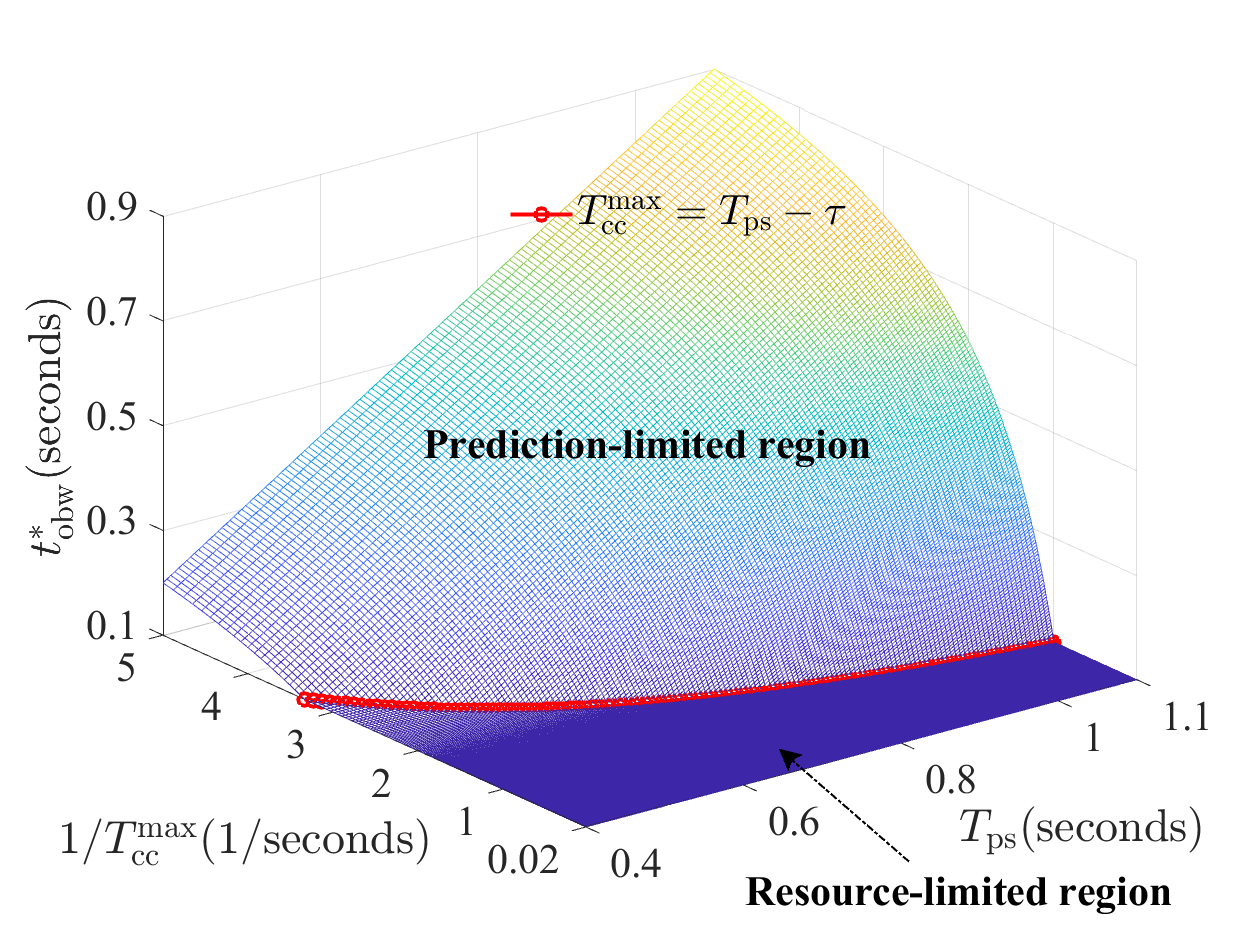}
        \end{minipage}
        \begin{minipage}[c]{0.75\linewidth}
            \centering
            \includegraphics[width=1\textwidth]{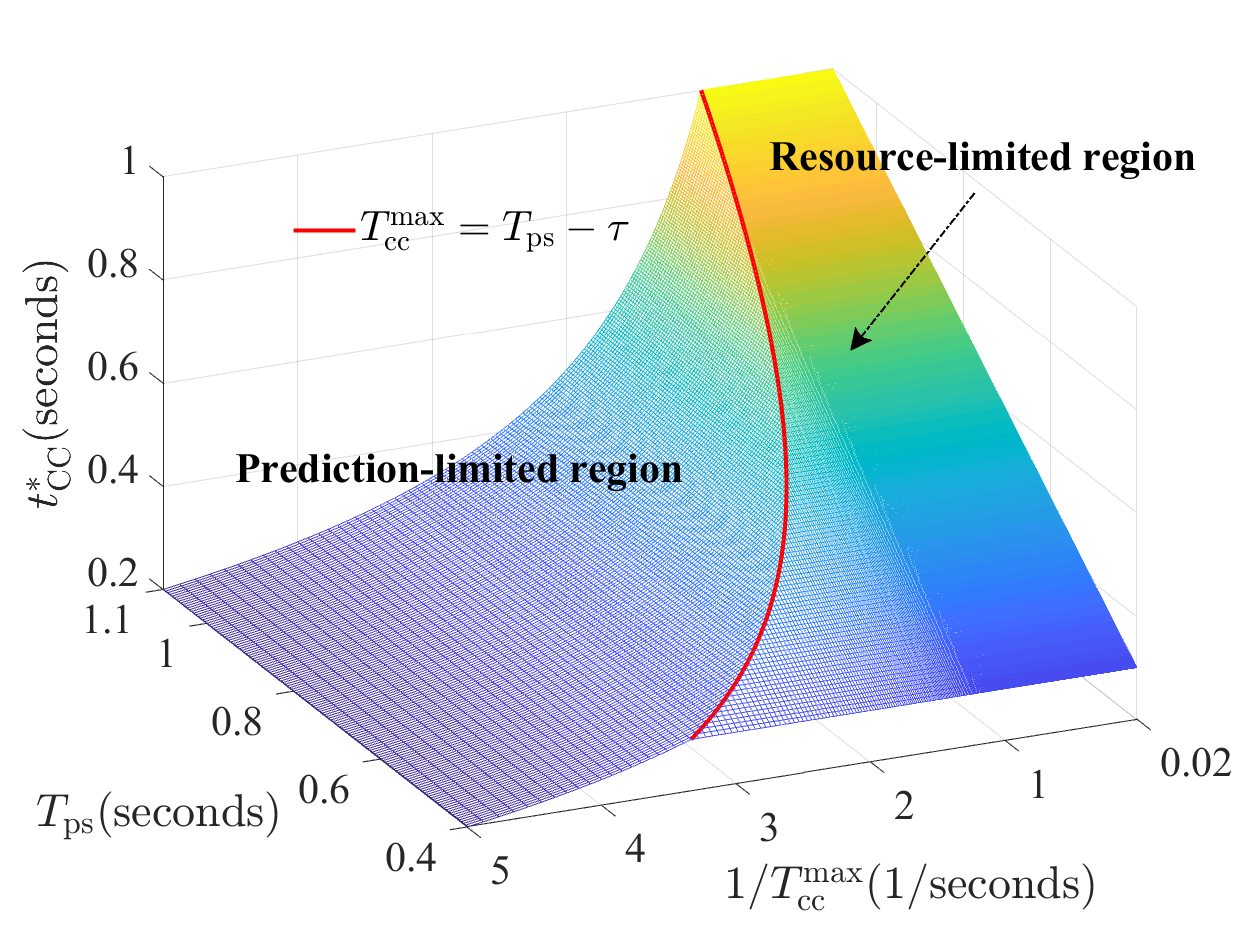}
        \end{minipage}
    \vspace{-0.3cm}
    \caption{Optimal durations v.s. $\frac{1}{T_{\mathrm{cc}}^{\max}}$ and $T_{\mathrm{ps}}$.}\label{Fig_structure of optimal solution}	
    \vspace{-0.6cm}
\end{figure}

To visualize the ``Resource-limited region" and ``Prediction-limited region", we provide the values of $t_{\mathrm{obw}}^{*}$ and $t_{\mathrm{cc}}^*$ obtained from \eqref{opt_slt_simplea} and \eqref{opt_slt_simpleb} given $\tau = 0.1$~s, respectively considering $T_{\mathrm{ps}} =  1$~s and showing the impact of different values of $T_{\mathrm{ps}}$ in Figs. \ref{Fig:opt_slt_2D_example} and \ref{Fig_structure of optimal solution}.

\vspace{-0mm}
\section{Simulation and Numerical Results}\label{simulation_result}
In this section, we first verify \textbf{Assumption 1} with three predictors, namely LR, CB, and Gated Recurrent Unit (GRU) neural network, by fitting their achieved prediction performance, average DoO, as the functions of $t_{\mathrm{obw}}$ over a real dataset. Then, we verify the feasible case of the solution with the three predictors and dataset via numerical results.
After that,
we show the QoE in the two regions achieved by the LR and CB methods, and demonstrate how the QoE is respectively improved by boosting the prediction performance and the completion rate of CC tasks with the CB method. Finally, we show the gain from balancing the three tasks in terms of improving QoE and the impact of the number of users.
Again, the solutions are all obtained from \eqref{opt_t_obw_general_case}, \eqref{opt_CC_general_case}, and \eqref{opt_S_CC_general_case} under the case that $\phi(H) > 0 \ \textrm{and}\ \Phi = \emptyset$.

The codes for reproducing all the results can be found at \cite{code}.

\vspace{-0mm}\subsection{$\mathcal{D}(t_{\mathrm{obw}})$ Functions under LR, CB and GRU Methods}\label{verifying_DoO}

The LR method in \cite{optimizing_VR} employs a head movement trace to train a linear model for predicting head movement, which then is mapped into the predicted tiles\cite{apcc}.
The CB method in \cite{apcc} predicts the tiles to be requested implicitly, by using the tile requests in an observation window as contexts and setting the tile requests in the next segment as arms.
The GRU method in \cite{LSTM_update} employs a sequence of tile requests in an observation window to predict the tile requests in the future FoVs in a segment where the segment contains only one frame, hence seg-DoO degradates to DoO.

We consider the real dataset in \cite{NTHU_dataset}, which contains 500 traces of tile requests from $K$ = 50 users watching 10 VR videos.
For the LR and CB methods, we make the prediction with the dataset and obtain the average seg-DoO. For the GRU method,
we use the DoO obtained in \cite{LSTM_update} that is also with the dataset in \cite{NTHU_dataset}.

Specifically, for the LR and CB method, $\tau = 0.1$~s\cite{apcc}, and we set $T_{\mathrm{ps}} = 1$~s.
For each video, $J$ = 50 traces are used to compute the ground-truth of the average seg-DoO. By making the prediction with each method using the $j$th trace as training set, $\mathcal{D}^j(t_{\mathrm{obw}})$ can be obtained under different durations of the observation window. Then, for each video, the ground-truth of the average seg-DoO can be obtained as $\mathcal{D}(t_{\mathrm{obw}}) = \frac{1}{J}\sum_{j=1}^{J}\mathcal{D}^j(t_{\mathrm{obw}})$, which is fitted in sequel.

For the LR method, the fitted function is
$\mathcal{D}_{\mathrm{LR}}(t_{\mathrm{obw}}) = \sum_{n=0}^{3} a_n t_{\mathrm{obw}}^n$,
where $\{a_n\}$ are the fitted coefficients.
For the CB method, the fitted function is
\begin{align}
\mathcal{D}_{\mathrm{CB}}(t_{\mathrm{obw}}) = a_1   t_{\mathrm{obw}} + a_0. \label{DoO_CB}
\end{align}
For the GRU method, $\tau = 1$~s and $T_{\mathrm{ps}} = 2$~s\cite{LSTM_update}. We take the DoO in \cite{LSTM_update} as the ground-truth of $\mathcal{D}(t_{\mathrm{obw}})$, for which the fitted function can be obtained as
$\mathcal{D}_{\mathrm{GRU}}(t_{\mathrm{obw}}) = \sum_{n=0}^{2} a_n t_{\mathrm{obw}}^n$.

To evaluate the fitting performance, we use mean square error (MSE) as the metric that is widely used in regression analysis, $\textrm{MSE} \triangleq \sum_{d=1}^{D} \big( \mathcal{D} (t_{\mathrm{obw},d}) - \widehat{\mathcal{D}}(t_{\mathrm{obw},d}) \big)^2$,
where $D$ is the number of values of $t_{\mathrm{obw}}$, $t_{\mathrm{obw},d}$ is the $d$th value of $t_{\mathrm{obw}}$, and  $\widehat{\mathcal{D}}(t_{\mathrm{obw},d})$ is the fitted value of $\mathcal{D}(t_{\mathrm{obw},d})$. For the LR and CB methods, $t_{\mathrm{obw}}$ is set within $[0.1,1]$~s, which is equally divided into $D = 28$ discrete values. 
For the GRU method, $t_{\mathrm{obw}}$ is set within $[1,2]$~s, which is equally divided into $D = 4$ discrete values \cite{LSTM_update}.

The corresponding fitting parameters for each video and the fitting performance are provided in Tables \ref{table:LR}, \ref{table:CMAB}, and \ref{table:GRU}, respectively.

\begin{table}[htbp]
\captionsetup{font={small}}
\vspace{-0.1cm}
\caption{LR fitting parameters and fitting performance}\label{table:LR}
\vspace{-0.1cm}
\begin{center}
\begin{tabular}{|c|c|c|c|c|c|}
\hline
\multirow{2}{*}{Video} & \multicolumn{4}{c|}{Fitting
 parameters} & \multirow{2}{*}{MSE} \\
\cline{2-5}
& $a_0$ & $a_1$ & $a_2$& $a_3$ &\\ \hline
1 & 0.7101&0.1594 &-0.1648& 0.0717 & 9.6782$\times$$10^{-7}$ \\ \hline
2 & 0.7274&0.1686 &-0.2354 &0.1215 & 7.7358$\times$$10^{-7}$  \\ \hline
3 & 0.5773&0.1462 &-0.1955 &0.1015 & 7.2133$\times$$10^{-7}$  \\ \hline
\textbf{4} & \textbf{0.7305}&\textbf{0.0759} &\textbf{-0.0879} &\textbf{0.0392} & \textbf{2.9301}$\mathbf{\times}$$\mathbf{10^{-7}}$  \\ \hline
5 & 0.7276&0.2081 &-0.2722 &0.1392 & 9.6005$\times$$10^{-7}$  \\ \hline
6 & 0.7155&0.05881 &-0.0090 &-0.0100 &7.0082$\times$$10^{-7}$  \\ \hline
7 & 0.7813&0.10309 &-0.1231 &0.0627 & 1.1473$\times$$10^{-6}$   \\ \hline
8 & 0.6883&0.04294 &-0.0484 &0.0178 & 9.9769$\times$$10^{-7}$  \\ \hline
9 & 0.7404&0.1499 &-0.1802 &0.0837 & 7.3371$\times$$10^{-7}$  \\ \hline
\textbf{10} & \textbf{0.7309}&\textbf{0.1060} &\textbf{-0.0816} &\textbf{0.0210} & \textbf{1.4887}$\times$$\mathbf{10^{-6}}$  \\ \hline
\end{tabular}
\end{center}
\vspace{-0.2cm}
\end{table}

\begin{table}[htbp]
\captionsetup{font={small}}
\vspace{-0.1cm}
\caption{CB fitting parameters and fitting performance}\label{table:CMAB}
\vspace{-0.1cm}
\begin{center}
\begin{tabular}{|c|c|c|c| }
\hline
\multirow{2}{*}{Video} & \multicolumn{2}{c|}{Fitting parameters} & \multirow{2}{*}{MSE}  \\
\cline{2-3}
& $a_0$ & $a_1$ & \\ \hline
1 &0.7702& 0.1242 & 4.7013$\times$$10^{-6}$ \\ \hline
\textbf{2} &\textbf{0.7665}& \textbf{0.1277} & \textbf{4.5542$\times$$\mathbf{10^{-6}}$}  \\ \hline
3 &0.6903& 0.1669 & 6.6648$\times$$10^{-6}$  \\ \hline
4 &0.7211 & 0.1544 & 8.1762$\times$$10^{-6}$  \\ \hline
5 &0.6578& 0.1833 & 7.5323$\times$$10^{-6}$  \\ \hline
\textbf{6} &\textbf{0.6680}& \textbf{0.1793} & \textbf{9.1398}$\times$$\mathbf{10^{-6}}$  \\ \hline
7 &0.7638& 0.1343 &5.2914$\times$$10^{-6}$   \\ \hline
8 &0.6565& 0.1863 & 9.0814$\times$$10^{-6}$  \\ \hline
9 &0.7183& 0.1562 & 7.1564$\times$$10^{-6}$  \\ \hline
10 &0.7008& 0.1648 & 8.4279$\times$$10^{-6}$  \\ \hline
\end{tabular}
\end{center}
\vspace{-0.2cm}
\end{table}

\begin{table}[htbp]
\vspace{-0.1cm}
\captionsetup{font={small}}
\caption{GRU fitting parameters and fitting performance}\label{table:GRU}
\vspace{-0.1cm}
\begin{center}
\begin{tabular}{|c|c|c|c|c|}
\hline
\multirow{2}{*}{Video} & \multicolumn{3}{c|}{Fitting
 parameters} & \multirow{2}{*}{MSE} \\
\cline{2-4}
& $a_0$ & $a_1$ & $a_2$&\\ \hline
1 & 0.8525&-0.4180 &0.1949 & 1.2312$\times$$10^{-5}$ \\ \hline
2 & 0.5764&0.0139 &0.0217  & 1.6985$\times$$10^{-4}$  \\ \hline
3 & 0.1242&0.2638 &-0.0253  & 3.2563$\times$$10^{-4}$  \\ \hline
\textbf{4} & \textbf{0.6032}&\textbf{-0.1540} &\textbf{0.1307} & \textbf{2.5126}$\mathbf{\times}$$\mathbf{10^{-7}}$  \\ \hline
5 & 0.1339&0.6118  &-0.1732 & 2.7362$\times$$10^{-4}$  \\ \hline
6 & 0.4739&0.1018  &0.0068 &2.7362$\times$$10^{-4}$  \\ \hline
\textbf{7} & \textbf{1.2081}&\textbf{-0.9462}  &\textbf{0.4012} & \textbf{4.2236}$\times$ $\mathbf{10^{-4}}$   \\ \hline
8 & 0.6508&-0.2979 &0.1755 & 1.2161$\times$$10^{-4}$  \\ \hline
9 & 0.4786&0.0.2221 &-0.0516  & 3.6181$\times$$10^{-5}$  \\ \hline
10 & -0.2761 &1.0700 &-0.2935 & 6.2814$\times$$10^{-6}$  \\ \hline
\end{tabular}
\end{center}
\vspace{-0.2cm}
\end{table}

\begin{figure}[htbp]
    \vspace{-0.8cm}
    \centering
    \subfigure[LR method, $T_{\mathrm{ps}} = 1$~s.]{\label{Fig:fitting_effect_LR}
        \begin{minipage}[c]{0.75\linewidth}
            \centering
            \includegraphics[width=1\textwidth]{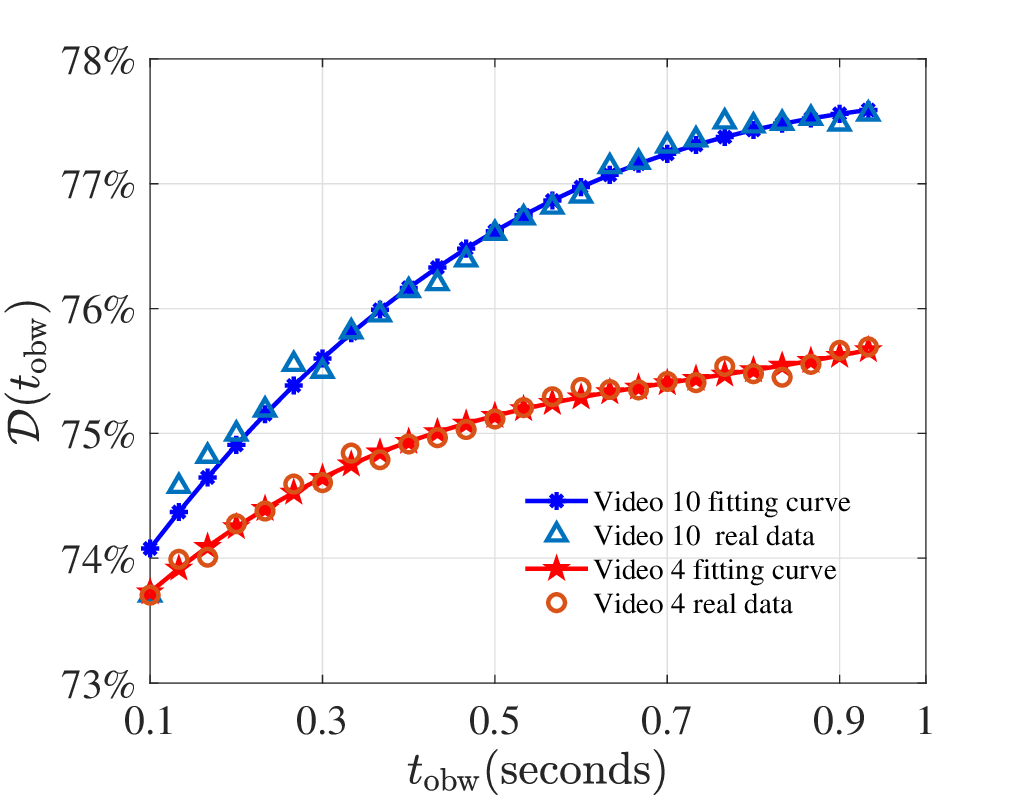}
        \end{minipage}
    }
    \subfigure[CB method, $T_{\mathrm{ps}} = 1$~s.]{\label{Fig:fitting_effect_CMAB}
        \begin{minipage}[c]{0.75\linewidth}
            \centering
            \includegraphics[width=1\textwidth]{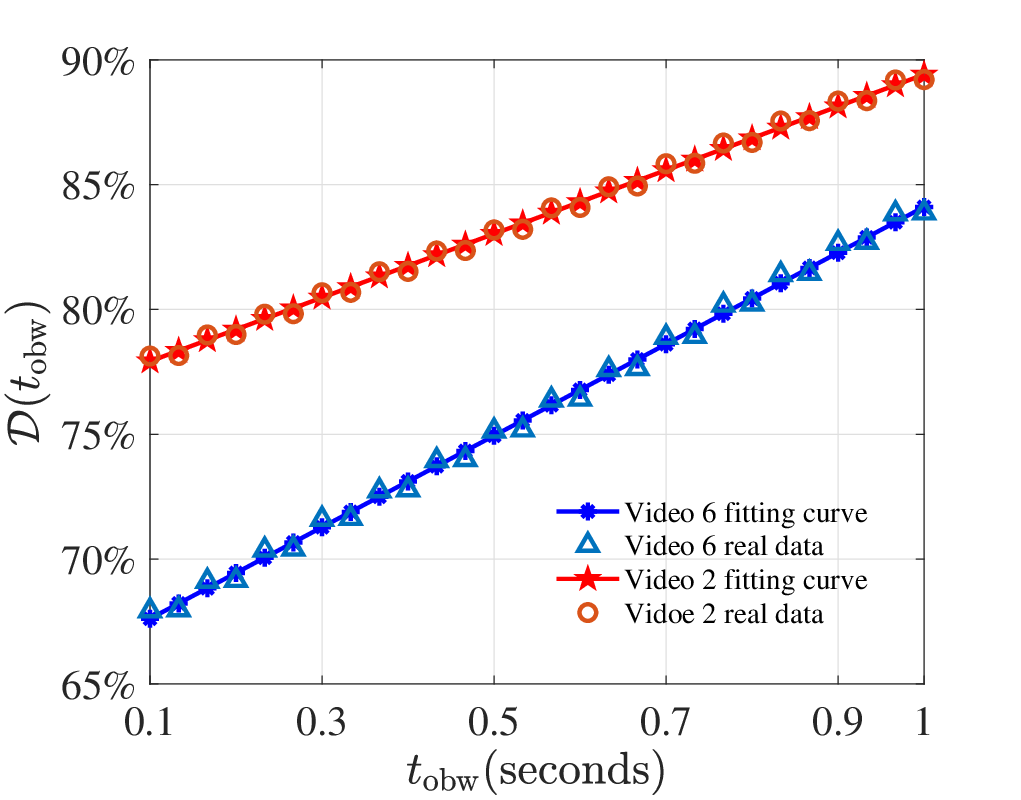}
        \end{minipage}
    }
    \subfigure[GRU method, $T_{\mathrm{ps}} = 2$~s.]{\label{Fig:fitting_effect_GrU}
        \begin{minipage}[c]{0.75\linewidth}
            \centering
            \includegraphics[width=1\textwidth]{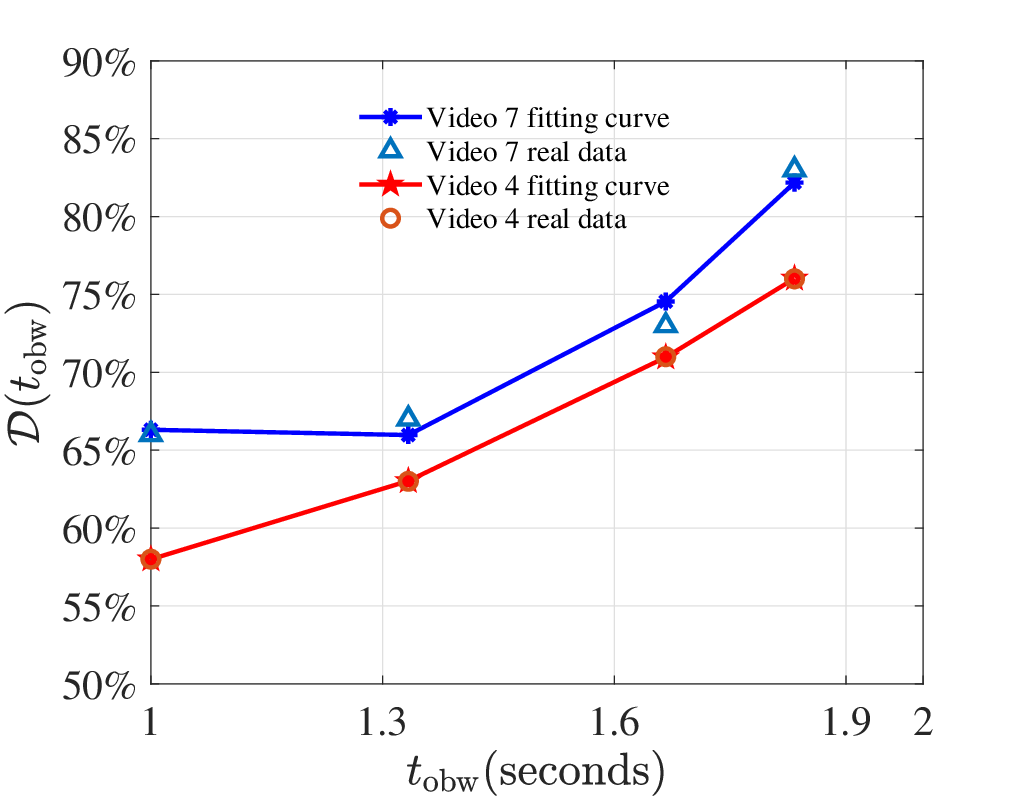}
        \end{minipage}
    }
    \vspace{-0.15cm}
    \caption{Prediction performance of three existing predictors and corresponding fitting functions.}\label{Examples_of_fitting_effects}	
    \vspace{-0.2cm}
\end{figure}

In Fig. \ref{Examples_of_fitting_effects}, we compare the fitting function and the prediction performance $\mathcal{D}(t_{\mathrm{obw}})$. For each method, two curves for the videos with the best and worst fitting performance (highlighted in the Tables)  are presented.
We can see that both the functions fit well, and the fitted functions for all methods on all videos are increasing functions of $t_{\mathrm{obw}}$, which verifies \textbf{Assumption 1}. Therefore, we use the fitted functions to obtain the prediction performance in the following.
\vspace{-0.3cm}

\subsection{Verify the Feasible Case of the Solution}
Now we show that the solutions in \eqref{opt_t_obw_general_case}, \eqref{opt_CC_general_case}, and \eqref{opt_S_CC_general_case} are feasible in the case with $\phi(H)>0$ and $\Phi = \emptyset$ for the considered three predictors and the real dataset.
By substituting $\mathcal{D}_{\mathrm{LR}}(t_{\mathrm{obw}})$, $\mathcal{D}_{\mathrm{CB}}(t_{\mathrm{obw}})$, and $\mathcal{D}_{\mathrm{GRU}}(t_{\mathrm{obw}})$ into \eqref{phi_def}, respectively, we obtain $\phi(t_{\mathrm{obw}})$ for the three predictors as
$\phi_{\mathrm{LR}}(t_{\mathrm{obw}}) = 4a_3t_{\mathrm{obw}}^3 + (3a_2 -3a_3 T_{\mathrm{ps}})t_{\mathrm{obw}}^2 + (2a_1 -2a_2 T_{\mathrm{ps}})t_{\mathrm{obw}} + a_0 - a_1 T_{\mathrm{ps}}$,
$\phi_{\mathrm{CB}}(t_{\mathrm{obw}}) = 2a_1 t_{\mathrm{obw}} + a_0 - a_1 T_{\mathrm{ps}}$, and
$\phi_{\mathrm{GRU}}(t_{\mathrm{obw}})= 3a_2 t_{\mathrm{obw}}^2 + (2a_1 -2a_2 T_{\mathrm{ps}})t_{\mathrm{obw}} + a_0 - a_1 T_{\mathrm{ps}}$.
Using the same values of $T_{\mathrm{ps}}$ and $\tau$ as in subsection \ref{verifying_DoO}, the numerical results with values of $T_{\mathrm{cc}}^{\max}$ ranging from 0 to 10 s show that $\phi(H)>0$ always holds.

Then, we can obtain the values of $t_{\mathrm{obw}}$ satisfying $\phi(t_{\mathrm{obw}})=0$ for the CB and GRU methods respectively as
$t_{\mathrm{obw},\mathrm{CB}} = \frac{T_{\mathrm{ps}}}{2} - \frac{a_0}{2a_1}$, $t_{\mathrm{obw},\mathrm{GRU},1} = \frac{2a_2T_{\mathrm{ps}} -2a_1 + \sqrt{ \Delta } }{6a_2}$, and $t_{\mathrm{obw},\mathrm{GRU},2} = \frac{2a_2T_{\mathrm{ps}} -2a_1 - \sqrt{ \Delta } }{6a_2}$,
where $\Delta\triangleq(2a_1 - 2a_2T_{\mathrm{ps}} )^2 - 12a_2(a_0 -a_1 T_{\mathrm{ps}})$.
The values of $t_{\mathrm{obw}}$ satisfying $\phi(t_{\mathrm{obw}})=0$  for the LR method can be obtained via Shengjing's formula from a cubic equation \cite{shengjin}.
Numerical results show that all these values are either less than zero or complex numbers, which violate the constraint $t_{\mathrm{obw}} \geq \max\{T_{\mathrm{ps}} - T_{\mathrm{cc}}^{\max},\tau\}>0$ in \eqref{obj_defe}. Therefore, $\Phi = \emptyset$ holds.

\vspace{-0mm}\subsection{Relation of ${T_{\mathrm{cc}}^{\max}}$ with Communication and Computing Resources}
${T_{\mathrm{cc}}^{\max}}$ in \eqref{T_CC_max} characterizes the tradeoff between computing rate (depending on the allocated computing resource, performance of computing units, and the video contents) and transmission rate (depending on the transmit power, bandwidth, the number of antennas, and propagation environment). To illustrate such a tradeoff, we provide numerical results under different computing and transmission setups in the following.

We first show how to obtain the parameter $\mu_r$ in the computing model in \eqref{C_cpt_final} using the measured values of $T_r$ and other parameters in literature.
We consider one video named ``Roller Coaster" in the dataset \cite{NTHU_dataset} that has been measured by different GPUs. The measured value of $T_r$ in \cite{edge_assisted} for the video is 47.13 ms, where the video  are rendered by Nvidia GTX970 GPU under Maxwell architecture (commercialized in 2015) with $\mathcal{C}_{r}$ = 3.92$\times10^{12}$ FLOPS, $R_w=4096$ and $R_h=2160$ \cite{edge_assisted}. The measured value of $T_r$ for the same video is 4.5 ms in \cite{wireless_reactive_VR}, which is rendered by Nvidia TITAN X GPU under Pascal architecture (commercialized in 2016) with $\mathcal{C}_{r} = 6.691\times10^{12}$ FLOPS, $R_w=2160$ and $R_h=1200$ \cite{wireless_reactive_VR}. For both results, $\gamma_{\textit{fov}} = 0.2$\cite{NTHU_dataset}, $b=12$ bits per pixel\cite{HuaWei_Cloud_VR}. Then, we can obtain $\mu_r$ as 1.04$\times10^5$ and 5.81$\times10^4$, respectively.

Since there are no measured results for rendering using the latest GPU with Turing architecture (commercialized in 2018) as in the NVIDA cloudXR solution\cite{Nvidia_cloudXR,Nvidia_MEC_parameter}, we roughly estimate $\mu_r$ from publicized measurement.
The gain in terms of reducing $\mu_r$ achieved by a new architecture can be obtained as $G_{\mu_r}\triangleq\frac{1.04\times10^5}{5.81\times10^4}=21.5$. By conservatively assuming that $\mu_r$ can be improved in the same gain \cite{GPU}, we can calculate the value of $\mu_r$ for Turing architecture as $\mu_r^T=(\bar{\mu}_r^M)/G_{\mu_r}^2 = 2.23\times10^3$ FLOPs/bit, where $\bar{\mu}_r^M$ represents value for the Maxwell architecture obtained by averaging over the 10 videos in \cite{edge_assisted}, considering that the values of $\mu_r$ differ for videos.

We consider the VR video with 4K resolution in 3840$\times$2160 pixels\cite{FoV_aware_tile}.
The playback duration of a segment is $T_{\mathrm{seg}}=1$~s\cite{FoV_aware_tile}. Each segment is divided into $M$ = 24 tiles in 4 rows and 6 columns, which can provide a good trade-off between encoding efficiency and bandwidth saving \cite{tile_number}. We set $\gamma_{\textit{fov}}=0.2$\cite{NTHU_dataset} and $\gamma_{\textit{extra}}=0.1$\cite{HuaWei_Cloud_VR}. Then, the number of predicted tiles is $N_{} = \left \lceil{\gamma_{\mathrm{pt}}M}\right \rceil=8$.
Each tile has the resolution of $r_w=3840/6=640$, and $r_h=2160/4=540$ pixels\cite{FoV_aware_tile}, the number of tiled-frame in a tile $N_{\textit{tf}}=30$\cite{NTHU_dataset}, $b=12$ bits per pixel\cite{HuaWei_Cloud_VR}. To ensure the QoE of VR users, consider that the tiles are compressed with lossless coding, where the compression ratio $\gamma_c=2.41$\cite{HEVC_lossless_coding}. Hence, the numbers of bits in each tile for transmission and for rendering are $s_{\mathrm{com}} = r_w r_h b  N_{\textit{tf}}/\gamma_{c} = 52$ Mbits and
$s_{\mathrm{cpt}} = r_w r_h bN_{\textit{tf}}=124$ Mbits, respectively.

When the distance between BS and users are identical to $d_k=5$~m or 201~m (considering the height of the HMD), the path loss models are $32.4+20\log_{10}(f_c)+17.3\log_{10}(d_k)$ in dB\cite{channel-standard} or $32.4+20\log_{10}(f_c)+30\log_{10}(d_k)$ in dB\cite{channel-standard}, the maximal transmit power are ${P}=24$ dBm\cite{channel-standard} (and $20$ dBm) or ${P}=53$~dBm\cite{mMIMOBS-setting} (and $46$~dBm), and consider Rician channel
 with the ratio between the line-of-sight (LoS) and non-LoS channel power as $10$~dB
  or Rayleigh channel.
The carrier frequency is $f_c = 3.5$ GHz, the noise spectral density is -174 dBm/Hz.
The ensemble-average rate $C_{\mathrm{com}}$ for each user is obtained by averaging over $10^6$ instantaneous transmission rates. The proactive streaming parameters are $T_{\mathrm{ps}}=1$~s and $\tau=0.1$~s, respectively. The numerical results are provided in Table \ref{table:indoor_numerical_results}, where cases (a), (b), (d), (e), (f) and (g) are in prediction-limited region, and cases (c) and (h) are in resource-limited region.

\begin{table*}[htbp]
\vspace{-0.1cm}
\captionsetup{font={small}}
\caption{Numerical examples of $T_{\mathrm{cc}}^{\max}$ under different settings, a GPU with Turing architecture is used}\label{table:indoor_numerical_results}
\vspace{-0.3cm}
\begin{center}
\begin{tabular}{|c|c|c|c|c|c|c|c|c|c|c|}
\hline
\multirow{2}{*}{Index}&\multirow{2}{*}{$K$}&\multirow{2}{*}{$d_k$(m)}&\multicolumn{4}{c|}{Communication Parameters}&\multicolumn{3}{c|}{Computing Parameters}&\multirow{2}{*}{$T_{\mathrm{cc}}^{\max}$(s)}\\
\cline{4-10}
&&&$N_t$& $\texttt{P}$ (dBm)& B (MHz)&$C_{\mathrm{com}}$ (Gbps)&GPU&$\mathcal{C}_{\mathrm{total}}$ (FLOPS)&$C_{\mathrm{cpt}}$ (Gbps)&\\ \hline
(a)&1&5 &4&20&40&0.88&T4&8.1$\times10^{12}$&4.3&0.70\\ \hline
(b)&2&5 &4&20&40&0.81&P40&$11.7\times10^{12}$&3.1&0.82\\ \hline
(c)&4&5 &8&20&40&0.78&P40&$11.7\times10^{12}$&1.6&1.16\\ \hline
(d)&4&5 &8&24&150&2.85&P40&$11.7\times10^{12}$&1.6&0.78\\ \hline
(e)&4&5 &8&24&80&1.59&RTX 8000&$16.3\times10^{12}$&2.2&0.71\\ \hline
(f)&1&201 &16&46&40&0.62&RTX 8000&$16.3\times10^{12}$&8.7&0.78\\ \hline
(g)&1&201 &64&53&40&0.79&T4&8.1$\times10^{12}$&4.3&0.75\\ \hline
(h)&10&201 &64&53&40&0.65&T4&8.1$\times10^{12}$&0.4&2.93\\ \hline
\end{tabular}
\end{center}
\vspace{-0.15cm}
\end{table*}

As shown in cases (f) and (g), doubling the computing resource is roughly equivalent to reducing the communication resource to one fourth.
As shown in cases (g) and (h), the system enters to the resource-limited region as $K$ increases.
By comparing cases (a) and (g), we can see that $T_{\mathrm{cc}}^{\max}$ is smaller for small distance when the available computing resources are  identical.

We have also computed $T_{\mathrm{cc}}^{\max}$ using a GPU with Maxwell (i.e., Nvidia GTX970) or Pascal architecture (i.e, Nvidia TITAN X) at the MEC server under these settings. The results show that the system always operates in the resource-limited region even for a single VR user.

\vspace{-0mm}\subsection{QoE in the Two Regions}\label{verify_QoE_region}
To understand how the QoE is improved by increasing the resources and by improving the prediction performance, we illustrate the average QoE in the resource-limited and prediction-limited regions achieved by the LR and CB methods.
We set $\tau = 0.1$~s and $T_{\mathrm{ps}} = 1$~s, while the results for other values of $T_{\mathrm{ps}}$ are similar.
The average QoE is the average value of the QoE obtained by substituting (\ref{opt_t_obw_general_case}) and (\ref{opt_CC_general_case}) into \eqref{QoE-Metric}, taken over the 10 videos.

\begin{figure}[htbp]
    \vspace{-0.5cm}
    \centering
    \subfigure[QoE achieved by two predictors.]{\label{Fig:QoEvs_T_CC_2D}
        \begin{minipage}[c]{0.75\linewidth}
            \centering
            \includegraphics[width=1\textwidth]{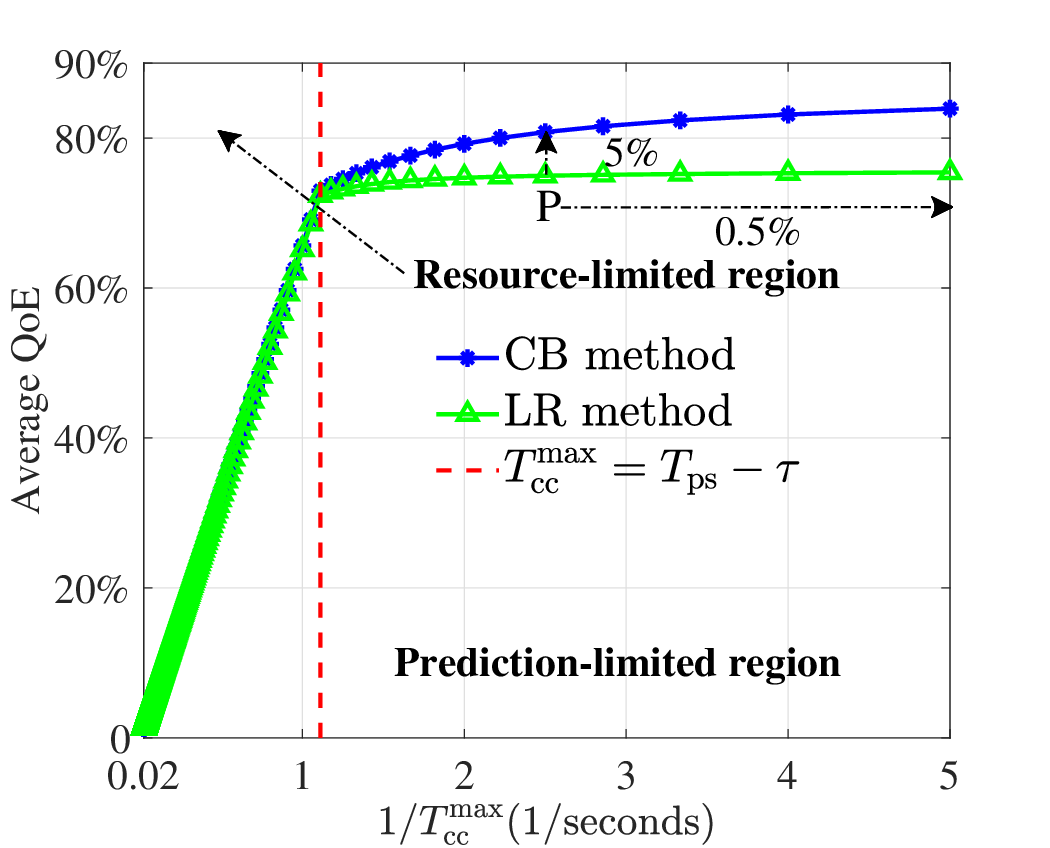}
        \end{minipage}
    }
    \subfigure[QoE, $\mathcal{D}(t_{\mathrm{obw}}^*)$, and $S_{\mathrm{cc}}^{*}(t_{\mathrm{cc}}^*)$  v.s. $T_{\mathrm{cc}}^{\max}$, CB method.]{\label{Fig:Matching_PCC_example_CB}
        \begin{minipage}[c]{0.75\linewidth}
            \centering
            \includegraphics[width=1\textwidth]{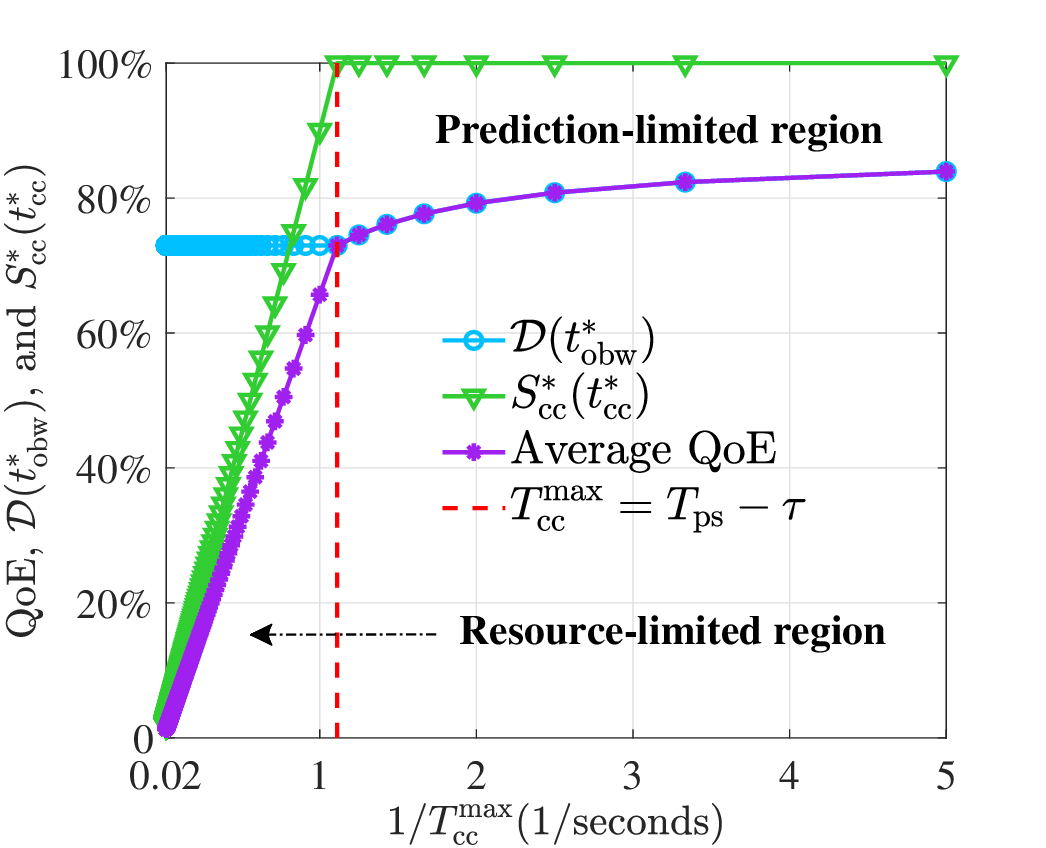}
        \end{minipage}
    }
    \vspace{-0.35cm}
    \caption{Impact of predictors, prediction performance and CC task completion rate on QoE.}\label{Examples_of_fitting_effect}	
    \vspace{-0.3cm}
\end{figure}



In Fig. \ref{Fig:QoEvs_T_CC_2D}, we show the average QoE achieved by two predictors versus the assigned resources to a user.
We observe that the average QoE increases rapidly in the resource-limited region, since the term $S_{\mathrm{cc}}(t_{\mathrm{com}},t_{\mathrm{cpt}})$ in the QoE metric grows with more resources.
By contrast, the average QoE  increases slowly in the prediction-limited region, because the increase of resources only plays a role of improving $\mathcal{D}(t_{\mathrm{obw}})$ with longer observation window by reducing the total duration for communication and computing.
In this region, the CB method outperforms the LR method in terms of the average QoE. This is because the CB method can achieve better prediction performance than the LR method, as shown in Fig. \ref{Fig:fitting_effect_LR} and Fig. \ref{Fig:fitting_effect_CMAB}.
Specifically, consider the average QoE achieved by the LR method when $1/{T_{\mathrm{cc}}^{\max}} = 2.5$, which is the point ``P'' in the figure. We can see that increasing the resources from $1/{T_{\mathrm{cc}}^{\max}} = 2.5$ to $1/{T_{\mathrm{cc}}^{\max}} = 5$
only improves the average QoE by 0.5\%. By contrast, when using the CB method for prediction, the average QoE is improved by 5\%. This indicates that using a better predictor provides much larger gain than using more resources in the prediction-limited region.

In Fig. \ref{Fig:Matching_PCC_example_CB}, we show how the average QoE is respectively enhanced by improving the prediction performance and by increasing the CC task completion rate. The value of $\mathcal{D}(t_{\mathrm{obw}}^*)$ is obtained by first substituting \eqref{opt_t_obw_general_case}
into \eqref{DoO_CB} and then being averaged over the 10 videos.
The value of   $S_{\mathrm{cc}}^{*}(t_{\mathrm{cc}}^*)$ is
obtained from \eqref{opt_S_CC_general_case}.
In the resource-limit region, the average QoE is improved by increasing the completion rate of CC tasks, while the prediction performance remains constant. In the prediction-limit region, the average QoE is improved by enhancing the prediction performance, while the completion rate of CC tasks remains constant.

In Table \ref{Table:example_cc_T_cc}, we provide two groups of transmission and computing rates to achieve identical value of $1/{T_{\mathrm{cc}}^{\max}}$, from which we illustrate how much communication resource or computing resource are required for further improving the QoE after the boundary line.


\begin{table*}[htbp]
\vspace{-0.1cm}
\captionsetup{font={small}}
\caption{Communication and computing rates to achieve three values of $1/{T_{\mathrm{cc}}^{\max}}$ in Fig. \ref{Examples_of_fitting_effect}}\label{Table:example_cc_T_cc}
\vspace{-0.3cm}
\begin{center}
\begin{tabular}{|c|c|c|c|c|}
\hline
\multirow{2}{*}{$1/T_{\mathrm{cc}}^{\max}$ (1/seconds)}&\multicolumn{2}{c|}{Group 1}&\multicolumn{2}{c|}{Group 2}\\
\cline{2-5}
&$C_{\mathrm{com}}$ (Gbps)& $C_{\mathrm{cpt}}$ (Gbps)& $C_{\mathrm{com}}$ (Gbps)&$C_{\mathrm{cpt}}$ (Gbps)\\ \hline
0.02&3.5&0.2&0.1&8.7\\ \hline
1.1 (Boundary line)&3.5&1.0&0.4&8.7\\ \hline
5&3.5&12.1&4.8&8.7\\ \hline
\end{tabular}
\end{center}
\end{table*}

\subsection{QoE Gain from Optimizing $t_{\mathrm{obw}}$ and $t_{\mathrm{cc}}$}\label{simulation_results}
We evaluate the performance gain from optimizing $t_{\mathrm{obw}}$ and $t_{\mathrm{cc}}$, which is with legend ``\textbf{Opt duration}", in terms of the average QoE taken over the 10 videos,
by comparing with the following three schemes without duration optimization. (1) $T_{\mathrm{cc}} = T_{\mathrm{obw}} = {T_{\mathrm{ps}}}/{2}$, which is with legend ``$\textbf{1:1 duration}$". (2) $T_{\mathrm{obw}} = {T_{\mathrm{ps}}}/{3}$, $T_{\mathrm{cc}} = {2T_{\mathrm{ps}}}/{3}$, i.e., more time is allocated for communication and computing, which is with legend ``$\textbf{1:2 duration}$". (3) $T_{\mathrm{obw}} = {2T_{\mathrm{ps}}}/{3}$, $T_{\mathrm{cc}} = {T_{\mathrm{ps}}}/{3}$, i.e., more time is allocated for prediction, which is with legend ``$\textbf{2:1 duration}$".

\begin{figure}[htbp]
\vspace{-0.3cm}
    \centering
    \subfigure[$T_{\mathrm{ps}} = 1$s, CB method.]{\label{Fig:opt_cmp_CB_T_ps1}
        \begin{minipage}[c]{0.75\linewidth}
            \centering
            \includegraphics[width=1\textwidth]{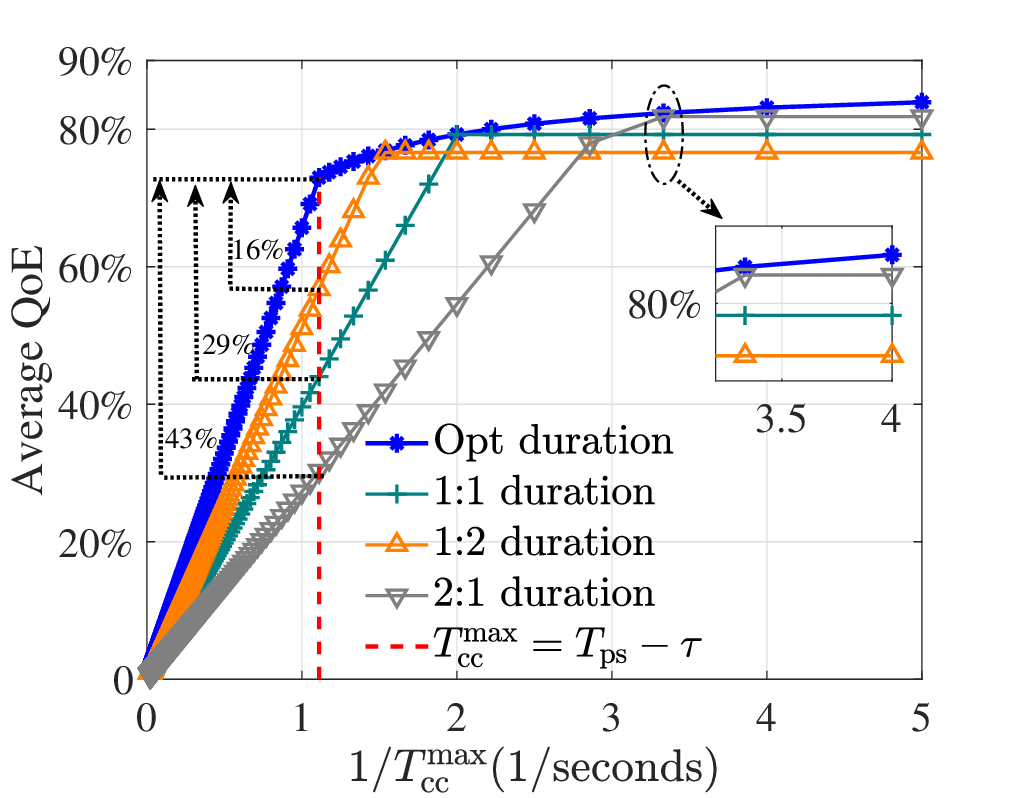}
        \end{minipage}
    }
    \subfigure[$T_{\mathrm{ps}} = 0.5$s, CB method.]{\label{Fig:opt_cmp_CB_T_ps05}
        \begin{minipage}[c]{0.75\linewidth}
            \centering
            \includegraphics[width=1\textwidth]{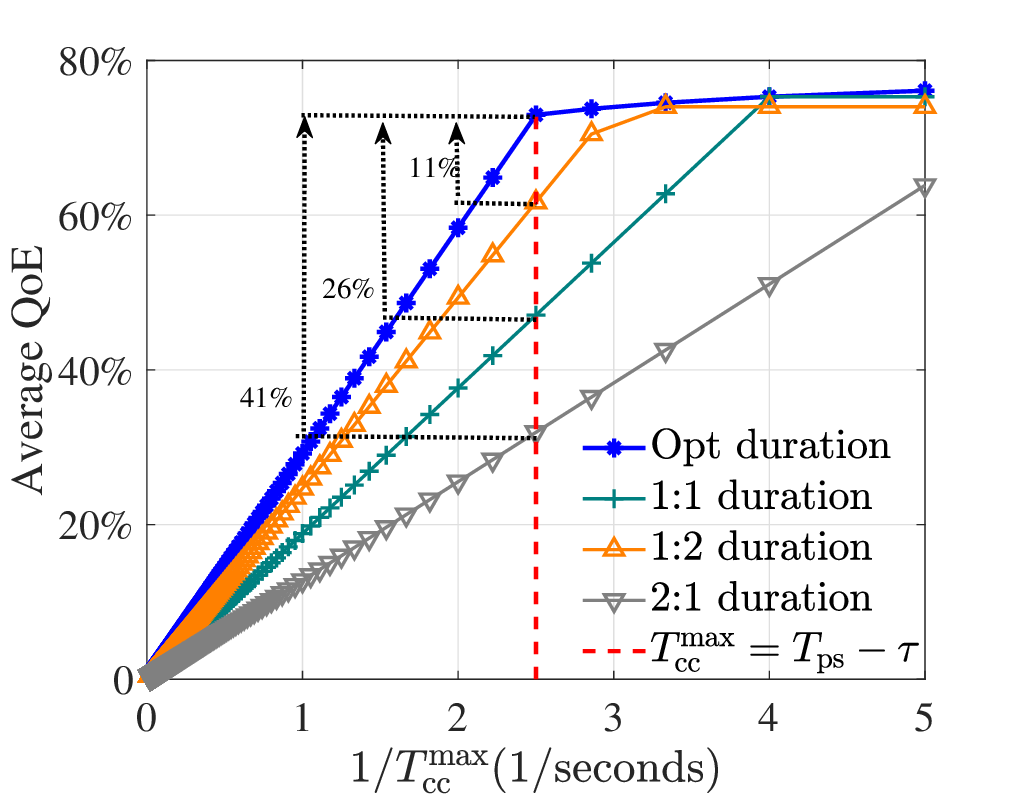}
        \end{minipage}
    }
    \centering
    \subfigure[$T_{\mathrm{ps}} = 1$s, LR method.]{\label{Fig:opt_cmp_LR_T_ps1}
        \begin{minipage}[c]{0.75\linewidth}
            \centering
            \includegraphics[width=1\textwidth]{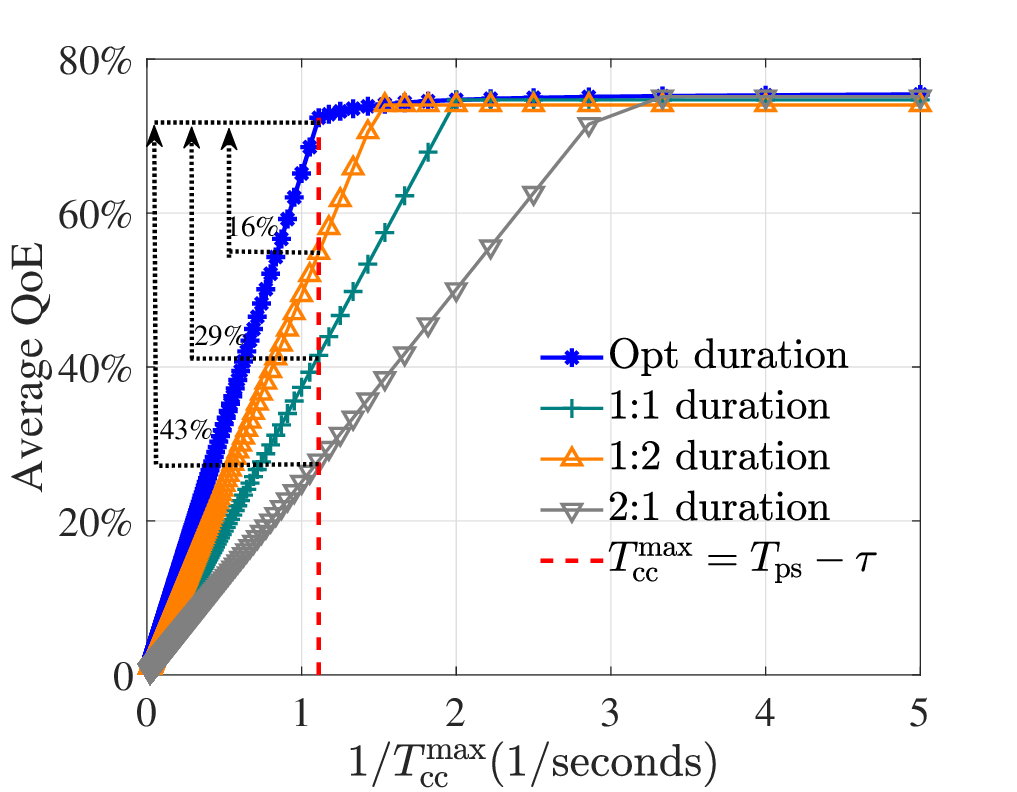}
        \end{minipage}
    }
    \subfigure[$T_{\mathrm{ps}} = 0.5$s, LR method.]{\label{Fig:opt_cmp_LR_T_ps05}
        \begin{minipage}[c]{0.75\linewidth}
            \centering
            \includegraphics[width=1\textwidth]{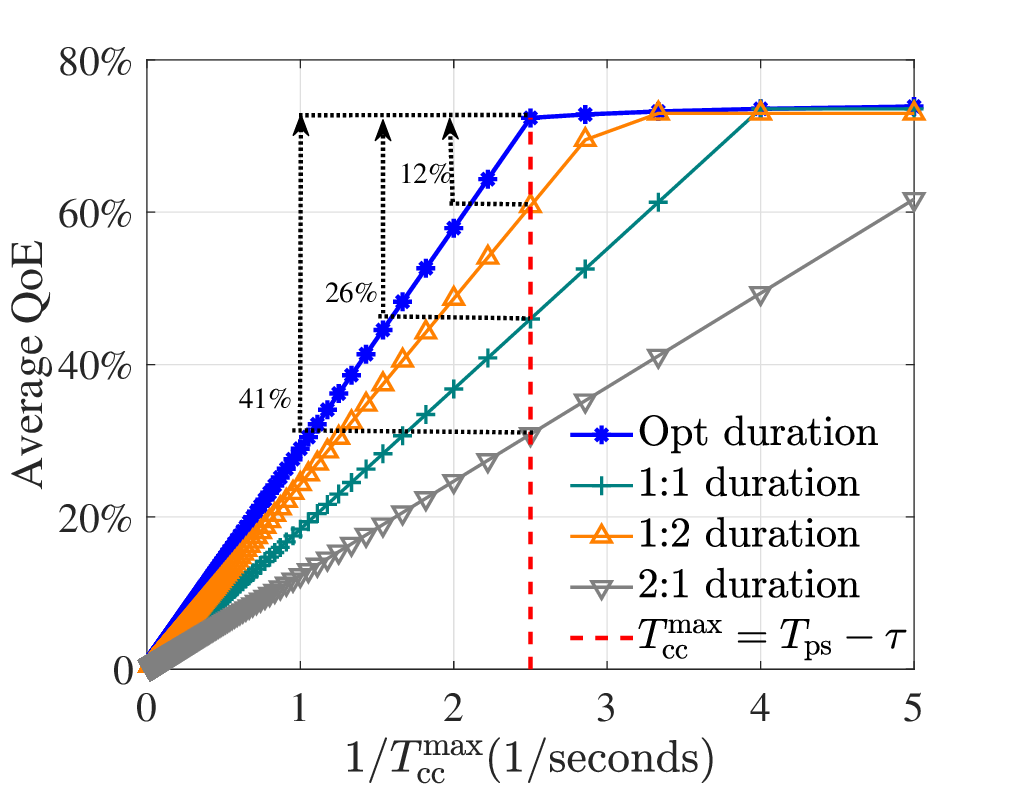}
        \end{minipage}
    }
    \vspace{-0.15cm}
    \caption{Performance comparison on average QoE.}\label{Fig:opt_cmp}	
    \vspace{-0.3cm}
\end{figure}

The QoE for each video is obtained by substituting (\ref{opt_t_obw_general_case}) and (\ref{opt_CC_general_case}) with the fitted function for each video into \eqref{P1_obj}. The average QoE achieved by the CB and LR methods are provided in Fig. \ref{Fig:opt_cmp}. As expected, the optimal solution always yields the best performance.
Specifically, we observe the following results.

\textbf{(1)} As the configured resources increase in the resource-limited region, the gain of the optimal solution over the three baselines becomes larger and achieves the maximum on the boundary line. As the resources further increase, the gain becomes smaller.
This is because in this region, the scheme with more time for communication and computing yields better performance. The optimal solution allocates largest value of $t_{\mathrm{cc}}$, thus can compute and deliver most predicted tiles among the four schemes.
When the configured resources achieve the boundary line, the optimal solution can compute and deliver all the predicted tiles and begins to enter into the prediction-limited region, while the three baselines are still in the resource-limited region.
The reason that the gain of the optimal solution becomes smaller after the boundary line is two-fold.
On the one hand, for the optimal solution,
the QoE increases slowly in the prediction-limited region.
On the other hand, the baselines are still in the resource-limited region, so that increasing the configured resources can increase the QoE rapidly.

\textbf{(2)} As the configured resources increase, the QoE achieved by the three non-optimized schemes first increases then remains constant, while the QoE achieved by the optimal solution still increases.  This is more clear in Fig. \ref{Fig:opt_cmp}\subref{Fig:opt_cmp_CB_T_ps1}. The reason that the QoE keeps constant for the baselines is that they are in the prediction-limited region but
still employ fixed duration for prediction. On the contrary, the optimal solution allocates more time to the observation window for assisting the prediction, 
thus the QoE can be further improved.

\textbf{(3)} For the optimal solution, the case with smaller $T_{\mathrm{ps}}$ shown in Fig. \ref{Fig:opt_cmp}\subref{Fig:opt_cmp_CB_T_ps05} and Fig. \ref{Fig:opt_cmp}\subref{Fig:opt_cmp_LR_T_ps05} needs more resources to achieve the boundary line, compared to the case with larger $T_{\mathrm{ps}}$ in Fig. \ref{Fig:opt_cmp}\subref{Fig:opt_cmp_CB_T_ps1} and Fig. \ref{Fig:opt_cmp}\subref{Fig:opt_cmp_LR_T_ps1}.
This is because when $T_{\mathrm{ps}}$  is smaller, $T_{\mathrm{cc}}^{\max}= T_{\mathrm{ps}} - \tau$ is smaller, and $1/{T_{\mathrm{cc}}^{\max}}$ is larger, i.e., more resources need to be configured to achieve the boundary line.

\textbf{(4)} For the average QoE, the gain of the CB method is higher than the gain of the LR method after the boundary line, because the CB method yields better prediction as shown in Fig. \ref{Fig:QoEvs_T_CC_2D}.

    \vspace{-0.1cm}

To show the impact of the number of the VR users on the QoE, we provide simulation with the same channel as in the 201 m cases of Table \ref{table:indoor_numerical_results} where the users are with identical distance.
The transmit power is ${P}=53$~dBm\cite{mMIMOBS-setting}, $N_t=64$, $B=100$ MHz.
A NVIDIA RTX 8000 GPU is used for rendering, where the computing resource $\mathcal{C}_{\mathrm{total}}=16.3\times10^{12}$ FLOPS is equally allocated among users. Then, $\frac{1}{T_{\mathrm{cc}}^{\max}}$ for each user is identical.

\begin{figure}[htbp]
\vspace{-0.5cm}
    \centering
        \begin{minipage}[c]{0.75\linewidth}
            \centering
            \includegraphics[width=1\textwidth]{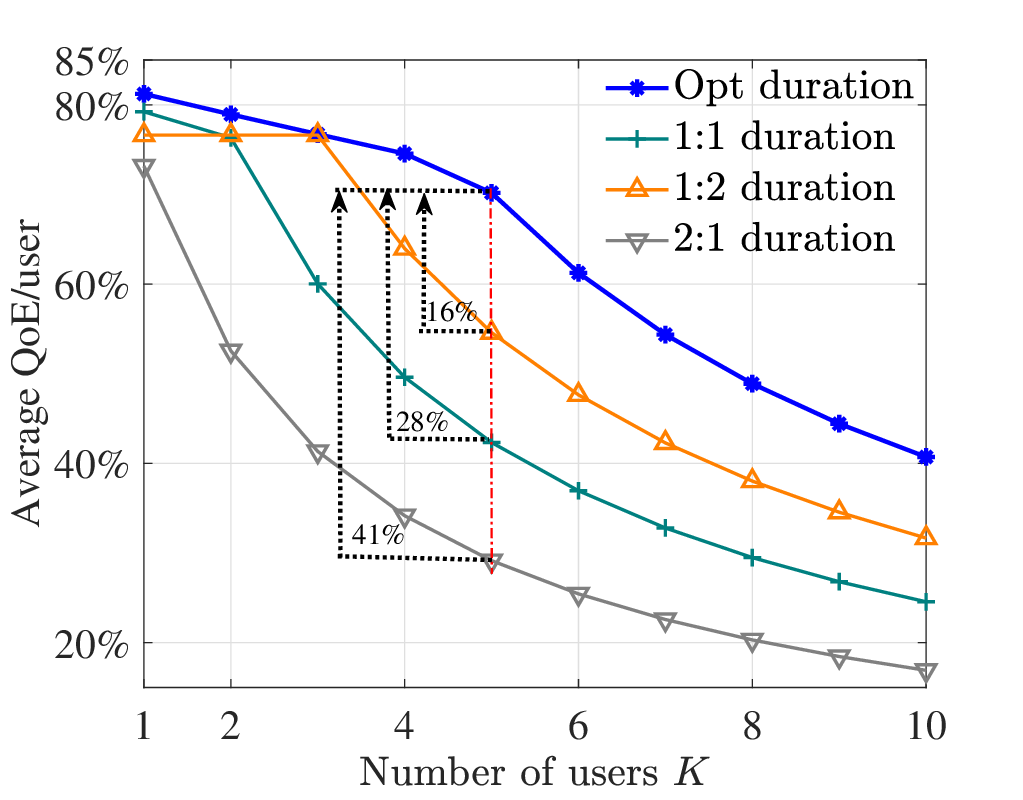}
        \end{minipage}
    \vspace{-0.15cm}
    \caption{Average QoE per user v.s. $K$, $T_{\mathrm{ps}}=1$~s and $\tau=0.1$~s, CB method}\label{Fig:multiuser_case}	
    \vspace{-0.3cm}
\end{figure}

As shown in Fig. \ref{Fig:multiuser_case}, the optimal solution always yields the best performance as expected. The achieved QoE first decreases slowly with $K$. This is because the optimal solution falls in the prediction-limited region when the number of users is small such that the resource reduction has little impact on the QoE, meanwhile the optimal solution can still exploit resources to improve the QoE even in this region though not significantly. As the resources further decreases, ``Opt duration'' enters the resource-limited region, and the QoE decreases rapidly.
The QoE of baseline ``1:2 duration'' first remains unchanged and then decreases with $K$. This is because when the number of users is small such that each user is allocated with sufficient resources, all the predicted tiles can be computed and transmitted with such durations, i.e., the baseline is in the prediction-limited region. In this region, the baseline
cannot exploit the extra resources for prediction, and hence the achieved QoE is not affected by the reduced resources. When $K$ is larger, the baseline ``1:2 duration'' enters the resource-limited region.
The QoE of other two baselines (i.e., ``1:1 duration" and ``2:1 duration'') decrease rapidly with $K$, because they lie in the resource-limited region.

\section{Conclusions}
In this paper, we studied how to match the communication and computing performance with the prediction performance to maximize the QoE of proactive tile-based VR streaming. To this end, we jointly optimized the durations for communication and computing with given transmission and computing rates and the duration of the observation window for prediction given any predictor.
With a reasonable assumption for watching one VR video, we obtained the closed-form optimal solution, from which we found a resource-limited region where the QoE can be remarkably improved by configuring more resources, and a prediction-limited region where designing better predictor is more effective in improving QoE.
Simulations with three existing tile predictors and a real dataset validated the employed assumption, and  demonstrated the performance gain from jointly optimizing durations for communication, computing and prediction.

\begin{appendices}
\numberwithin{equation}{section}
\section{Proof of the Solution of Problem \eqref{P3}}\label{Proof:MEC_t_com_t_cpt_KKT}
The Karush-Kuhn-Tucker (KKT) conditions of problem \eqref{P3} can be expressed as
\begin{subequations}\label{P3_KKT}
\begin{align}
  \frac{C_{\mathrm{com}}}{s_{\mathrm{com}}N} \lambda_1 - \frac{C_{\mathrm{cpt}}}{s_{\mathrm{cpt}}N}\lambda_2 = 0, \label{P3_KKT_lambda_1_lambda_2}\\
  -1 + \lambda_1 + \lambda_2 + \lambda_3 = 0, \label{P3_KKT_lambda_1_2_3}\\
  \lambda_1 \big(S_{\mathrm{cc}}(t_{\mathrm{cc}}) - \frac{C_{\mathrm{com}}(t_{\mathrm{cc}} - t_{\mathrm{cpt}}(t_{\mathrm{cc}}))}{s_{\mathrm{com}}N}\big) = 0,\label{P3_KKT_constraint_lambda_1}\\
\lambda_2 \big(S_{\mathrm{cc}}(t_{\mathrm{cc}}) - \frac{C_{\mathrm{cpt}}t_{\mathrm{cpt}}(t_{\mathrm{cc}})}{s_{\mathrm{cpt}}N}\big) = 0 ,\label{P3_KKT_constraint_lambda_2}\\
\lambda_3 \big(S_{\mathrm{cc}}(t_{\mathrm{cc}}) - 1\big) = 0,\label{P3_KKT_constraint_lambda_3}\\
  \eqref{P3_C1}-\eqref{P3_C3}, \lambda_1,\lambda_2,\lambda_3\geq 0.
\end{align}
\end{subequations}

From \eqref{P3_KKT_lambda_1_lambda_2}, we obtain that both $\lambda_1$ and $\lambda_2$ are either positive or equal to zero, thus there are two cases.

When $\lambda_1, \lambda_2>0$,
from \eqref{P3_KKT_constraint_lambda_1} and \eqref{P3_KKT_constraint_lambda_2} we have
\begin{align}\label{P3_KKT_S_CC}
S_{\mathrm{cc}}(t_{\mathrm{cc}}) =  \frac{C_{\mathrm{com}}\big(t_{\mathrm{cc}} - t_{\mathrm{cpt}}(t_{\mathrm{cc}})\big)}{s_{\mathrm{com}}N} = \frac{C_{\mathrm{cpt}}t_{\mathrm{cpt}}(t_{\mathrm{cc}})}{s_{\mathrm{cpt}}N}.
\end{align}
We can obtain $t_{\mathrm{cpt}}(t_{\mathrm{cc}})$ from \eqref{P3_KKT_S_CC} as $t_{\mathrm{cpt}}(t_{\mathrm{cc}}) = \frac{C_{\mathrm{com}}s_{\mathrm{cpt}}N}{C_{\mathrm{com}}s_{\mathrm{cpt}}N + C_{\mathrm{cpt}}s_{\mathrm{com}}N} t_{\mathrm{cc}}$.
Upon substituting into \eqref{P3_KKT_S_CC}, we have $S_{\mathrm{cc}}(t_{\mathrm{cc}}) = \frac{C_{\mathrm{com}} C_{\mathrm{cpt}}}{C_{\mathrm{com}}s_{\mathrm{cpt}}N + C_{\mathrm{cpt}}s_{\mathrm{com}}N} t_{\mathrm{cc}}.$
Upon substituting into
\eqref{P3_C3}, we have $ \frac{C_{\mathrm{com}} C_{\mathrm{cpt}}}{C_{\mathrm{com}}s_{\mathrm{cpt}}N + C_{\mathrm{cpt}}s_{\mathrm{com}}N} t_{\mathrm{cc}}\leq 1$, and after some regular derivations, we obtain $t_{\mathrm{cc}} \leq \frac{s_{\mathrm{com}}N}{C_{\mathrm{com}}} + \frac{s_{\mathrm{cpt}}N}{C_{\mathrm{cpt}}}$.

When $\lambda_1, \lambda_2 = 0$
from \eqref{P3_KKT_lambda_1_2_3}, we have $\lambda_3 = 1>0$. From \eqref{P3_KKT_constraint_lambda_3}, we have $S_{\mathrm{cc}}(t_{\mathrm{cc}}) = 1$. Upon substituting into \eqref{P3_C2} and \eqref{P3_C1}, we obtain
\begin{subequations}
\begin{align}
t_{\mathrm{cpt}}(t_{\mathrm{cc}})&\geq \frac{s_{\mathrm{cpt}}N}{C_{\mathrm{cpt}}},\label{P3_KKT_t_cpt>threshold}\\
t_{\mathrm{cc}} &\geq \frac{s_{\mathrm{com}}N}{C_{\mathrm{com}}} + t_{\mathrm{cpt}}(t_{\mathrm{cc}}).\label{P3_KKT_t_cc>threshold}
\end{align}
\end{subequations}
%
By substituting \eqref{P3_KKT_t_cpt>threshold} into \eqref{P3_KKT_t_cc>threshold}, we obtain $t_{\mathrm{cc}} \geq \frac{s_{\mathrm{com}}N}{C_{\mathrm{com}}} + \frac{s_{\mathrm{cpt}}N}{C_{\mathrm{cpt}}}$.
From these two cases, the solution of problem \eqref{P3} can be obtained.

\section{ Proof of the Solution of Problem \eqref{P4} }\label{Proof:MEC_t_obw_KKT}

By eliminating the variable $t_{\mathrm{cc}} =  T_{\mathrm{ps}} - t_{\mathrm{obw}}$,
problem \eqref{P4} becomes
\begin{subequations}\label{P5}
\begin{align}
  &  \min_{t_{\mathrm{obw}}} \   \ \   -\mathcal{D}(t_{\mathrm{obw}})\frac{(T_{\mathrm{ps}} - t_{\mathrm{obw}} )}{T_{\mathrm{cc}}^{\max}}  \label{P5_obj}\\
  & \ \ s.t.\ \ \    T_{\mathrm{ps}} - t_{\mathrm{obw}} \leq T_{\mathrm{cc}}^{\max},\label{P5_C1}\\
  & \ \ \ \ \ \  \ \ \   -t_{\mathrm{obw}} \leq -\tau.\label{P5_C2}
\end{align}
\end{subequations}

The KKT conditions of problem \eqref{P5} can be expressed as
\begin{subequations}
\begin{align}
  &D^{'}(t_{\mathrm{obw}})(T_{\mathrm{ps}} - t_{\mathrm{obw}} )  - D(t_{\mathrm{obw}}) + \lambda_1 + \lambda_2 = 0,\label{P5_KKt_K1}\\
  &\lambda_1(T_{\mathrm{ps}} - t_{\mathrm{obw}} - T_{\mathrm{cc}}^{\max}) = 0, \label{P5_KKt_K2}\\
  &\lambda_2(\tau - t_{\mathrm{obw}}) = 0,\label{P5_KKt_K3}\\
  & \eqref{P5_C1},\eqref{P5_C2}, \lambda_1,\lambda_2 \geq 0,
\end{align}
\end{subequations}
where $\mathcal{D}^{'}(t_{\mathrm{obw}})$ is the derivative function of $\mathcal{D}(t_{\mathrm{obw}})$.

There are four cases in the KKT conditions.
(1) When $\lambda_1 > 0, \lambda_2 = 0$, we have $t_{\mathrm{obw}} = T_{\mathrm{ps}} - T_{\mathrm{cc}}^{\max} $ from \eqref{P5_KKt_K2}, $T_{\mathrm{ps}} - T_{\mathrm{cc}}^{\max}\geq \tau$ from \eqref{P5_C2}, and we obtain $\phi(t_{\mathrm{obw}}) > 0 $ from \eqref{P5_KKt_K1} where $\phi(t_{\mathrm{obw}}) = \sum_{n=0}^{\infty} a_n t_{\mathrm{obw}}^n - (\sum_{n=1}^{\infty} n a_n t_{\mathrm{obw}}^{n-1})(T_{\mathrm{ps}} - t_{\mathrm{obw}})$.
(2) When $\lambda_1 = 0, \lambda_2 > 0$, we have $t_{\mathrm{obw}} = \tau$  from \eqref{P5_KKt_K3}, $\tau \geq T_{\mathrm{ps}} - T_{\mathrm{cc}}^{\max}$ from \eqref{P5_C1}, and $\phi(t_{\mathrm{obw}}) > 0 $ from \eqref{P5_KKt_K1}.
(3) When $\lambda_1 > 0, \lambda_2 > 0$, we have $t_{\mathrm{obw}} = T_{\mathrm{ps}} - T_{\mathrm{cc}}^{\max}$  from \eqref{P5_KKt_K2}, $T_{\mathrm{ps}} - T_{\mathrm{cc}}^{\max} = \tau$ from \eqref{P5_C2}, and $\phi(t_{\mathrm{obw}}) > 0 $ from \eqref{P5_KKt_K1}.
(4) When $\lambda_1 = 0, \lambda_2 = 0$, we obtain $t_{\mathrm{obw}} \geq \max\{T_{\mathrm{ps}} - T_{\mathrm{cc}}^{\max}, \tau \}$ from \eqref{P5_C1} and \eqref{P5_C2}, and $\phi(t_{\mathrm{obw}}) = 0$ from \eqref{P5_KKt_K1}.
Since the monotonicity of $\phi(t_{\mathrm{obw}})$ is unknown, there are three cases in the solutions of $\phi(t_{\mathrm{obw}}) = 0$: i) multiple different values,
ii) only one value,
iii) no real number value.
Besides, the solutions of $\phi(t_{\mathrm{obw}}) = 0$ should satisfy $t_{\mathrm{obw}} \geq \max\{T_{\mathrm{ps}} - T_{\mathrm{cc}}^{\max}, \tau \}$.
In the above feasible solutions, we find the optimal solution that maximizes the objective function in \eqref{P5_obj}. Thus we consider the following problem
\begin{subequations}\label{P_phi_opt}
\begin{align}
  &  \max_{t_{\mathrm{obw}}} \   \ \   f(t_{\mathrm{obw}})\triangleq\mathcal{D}(t_{\mathrm{obw}})\frac{(T_{\mathrm{ps}} - t_{\mathrm{obw}} )}{T_{\mathrm{cc}}^{\max}}  \\
  & \ \ s.t.\ \ \    \phi(t_{\mathrm{obw}}) = 0,\\
  & \ \ \ \ \ \  \ \ \   t_{\mathrm{obw}} \geq \max\{T_{\mathrm{ps}} - T_{\mathrm{cc}}^{\max}, \tau \}.
\end{align}
\end{subequations}
The feasible set of problem \eqref{P_phi_opt}  can be expressed as
\begin{align}\label{Phi_big_def}
\Phi\triangleq\{t_{\mathrm{obw}}|\phi(t_{\mathrm{obw}}) = 0, t_{\mathrm{obw}} \geq \max\{T_{\mathrm{ps}} - T_{\mathrm{cc}}^{\max}, \tau \}\}.
\end{align}
When $\Phi$ is not an empty set, i.e., at least one value of $t_{\mathrm{obw}}$ can satisfy $\phi(t_{\mathrm{obw}}) = 0$ and $t_{\mathrm{obw}} \geq \max\{T_{\mathrm{ps}} - T_{\mathrm{cc}}^{\max}, \tau \}$, the optimal solution can be obtained as
\begin{align}\label{T_phi}
T_{\Phi}^{*} = \mathop{\mathrm{argmax}}\limits_{t_{\mathrm{obw}}} f(t_{\mathrm{obw}}), t_{\mathrm{obw}}\in\Phi,
\end{align}
by exhaustive searching. Since the number of feasible solutions (i.e., $|\Phi|$) is finite, the complexity of searching is acceptable.
When $\Phi$ is an empty set, no feasible solution in this case.

From the first three cases, we can obtain that
\begin{align}\label{t_obw_H}
t_{\mathrm{obw}} = H = \max\{T_{\mathrm{ps}} - T_{\mathrm{cc}}^{\max},\tau\}, \ \textrm{if}\ \phi(H) >0.
\end{align}

From the fourth case, we can obtain that
$t_{\mathrm{obw}} = T_{\Phi}^*, \ \textrm{if}\  \Phi\neq \emptyset$.

When only one of the two cases holds, i.e.,  $\phi(H) >0$ and $\Phi = \emptyset$, or $\phi(H) = 0$ and $\Phi\neq \emptyset$, the solution 
is
\begin{align}\label{opt_slt:t_obw_1}
t_{\mathrm{obw}}^{*}=
\left
\{\begin{array}{ll}
H ,&  \phi(H) > 0 \ \textrm{and}\ \Phi = \emptyset,   \\
T_{\Phi}^*,   & \phi(H) = 0 \ \textrm{and}\ \Phi \neq \emptyset.
\end{array}
\right.
\end{align}

When both cases hold, i.e., $\phi(H) >0$ and $\Phi\neq \emptyset$, we can obtain the solution by comparing the objective function in the two cases.
Then, the solution is
\begin{align}\label{opt_slt:t_obw_2}
t_{\mathrm{obw}}^{*}=
\left
\{\begin{array}{ll}
H ,&  \phi(H) > 0,  \Phi \neq \emptyset \ \textrm{and}\  f(H) \geq f(T_{\Phi}^*),\\
T_{\Phi}^*,   & \phi(H) > 0,  \Phi \neq \emptyset \ \textrm{and}\ f(H) < f(T_{\Phi}^*).
\end{array}
\right.
\end{align}

The case that $\phi(H) = 0$ and $\Phi = \emptyset$ does not exist. This is because by substituting $\phi(H) = 0$ into \eqref{Phi_big_def}, at least one value $t_{\mathrm{obw}} = H$ can be found in $\Phi$. Thus, $\Phi \neq \emptyset$, and this case can be omitted.

From \eqref{opt_slt:t_obw_1} and \eqref{opt_slt:t_obw_2}, the solution of $t_{\mathrm{obw}}^{*}$ can be obtained as in \eqref{opt_t_obw_general_case}.
By substituting $t_{\mathrm{obw}}^{*}$ into $t_{\mathrm{cc}} =  T_{\mathrm{ps}} - t_{\mathrm{obw}}$, the solution of $t_{\mathrm{cc}}^*$ can be obtained as \eqref{opt_CC_general_case}.

\end{appendices}

\bibliographystyle{IEEEtran}
\bibliography{IEEEabrv,ref}

\end{document}